\documentclass[twocolumn]{emulateapj}
\defcitealias{RSR02}{RSR}
\defcitealias{RS01}{RS}

\begin{document} 

\title{The Impact of Galaxy Cluster Mergers on Cosmological 
Parameter Estimation from Surveys of the Sunyaev-Zel'dovich Effect}

\author{Daniel R. Wik and Craig L. Sarazin}
\affil{Department of Astronomy, University of Virginia,
       P.O. Box 400325, Charlottesville, VA 22904-4325;
       drw2x@virginia.edu, cls7i@virginia.edu}

\author{Paul M. Ricker}
\affil{Department of Astronomy, University of Illinois, 
       1002 W. Green St., Urbana, IL 61801; pmricker@uiuc.edu}

\and

\author{Scott W. Randall}
\affil{Harvard-Smithsonian Center for Astrophysics, 60 Garden St., 
       Cambridge, MA, 02138; srandall@cfa.harvard.edu}

\begin{abstract}
Sensitive surveys of the Cosmic Microwave Background (CMB)
will detect thousands of galaxy clusters via the Sunyaev-Zel'dovich 
(SZ) effect.
Two SZ observables, the central or maximum and integrated Comptonization
parameters $y_{\rm max}$ and $Y$, relate in a simple 
way to the total cluster mass, which allow the construction 
of mass functions (MFs) that can be used to estimate cosmological 
parameters such as the ratio of the
average matter density to the critical density $\Omega_M$, the
normalization of the spectrum of initial density perturbations
$\sigma_8$, and the dark energy equation of state parameter $w$.
However, clusters form from the mergers of smaller structures,
events that can disrupt
the equilibrium of intracluster gas upon which SZ--$M$
relations rely.
From a set of N-body/hydrodynamical simulations of binary cluster
mergers, we calculate the evolution of $Y$ and $y_{\rm max}$ over the
course of merger events and find that both parameters are transiently 
``boosted," primarily during the first core passage.
We then use a semi-analytic technique developed by \citet{RSR02}
to estimate the effect of merger boosts on the distribution functions 
YF and yF of $Y$ and $y_{\rm max}$, respectively,
via cluster merger histories
determined from extended Press-Schechter (PS) merger trees.
The scatter in the $Y$--$M$ and $y_{\rm max}$--$M$ relations from merger 
boosts are found to
be $\sim2\%$ and 25--30\% respectively.
To determine $\Omega_M$, $\sigma_8$, and $w$,
the boosted and nonboosted YFs and yFs are fit with analytic PS
distributions as a function of redshift.
We find that boosts do not induce an overall systematic effect on 
YFs, and the values of $\Omega_M$, $\sigma_8$, and $w$ (assumed constant)
were returned to within 2\% of values expected from the nonboosted YFs.
The boosted yFs are significantly biased, however, causing $\Omega_M$
to be underestimated by 15-45\%, $\sigma_8$ to be overestimated by
10-25\%, and $w$ to be pushed to more negative values by 25-45\%.
We also fit YF as a function of redshift to cosmological models in
which the dark energy parameter $w$ varied with redshift to assess the
effects of mergers on the inferred change in $w$ with redshift.
The values of $\Omega_M$, $\sigma_8$, and the low-redshift value
of $w$ ($w_0$) were again reproduced to within 2\%.
For the largest change in $w$ with $z$, which occurred between $z=0$ and
$z=1$ for the models assumed, it was increased by about 0.04.
Although this is twice as large as the merger effect on a constant value
of $w$, it is still reasonably modest.
We confirm that the integrated SZ effect, $Y$, is far more
robust to mergers than $y_{\rm max}$, as previously reported by 
\citet{MHB+05} and similarly found for the X-ray equivalent $Y_X$
\citep{KVN06, PBM+07},
and we conclude that $Y$ is the superior choice for a mass proxy when using SZ
observations of galaxy clusters to constrain cosmological parameters.
\end{abstract}


\section{Introduction} \label{sec:intro}
The evolution of galaxy cluster abundance traces 
the massive end of the
spectrum of
initial density fluctuations
and therefore is sensitive to cosmological parameters such as the
ratio of the average matter density to the critical density 
$\Omega_M \equiv 8\pi G \overline \rho /(3 H_0^2)$, 
the normalization of the power spectrum of initial density fluctuations 
$\sigma_8$, and the dark energy equation of state parameter 
$w$, equal to the ratio of the pressure to the energy density of
dark energy.
Here, $H_0$ is the Hubble constant and $\overline \rho$ is the average
density in the universe.
This sensitivity exists due to an exponential turnover at high
masses in the mass function (MF) of clusters, which can be predicted
from a well-established theoretical framework 
\citep[e.g.][]{HA91, KS96, HMH01}. 
However, only gravitational lensing, 
which remains observationally challenging,
directly measures the total mass of clusters.
In order to get masses for the large number of clusters needed to
construct the MF, it is often necessary to use 
a more observationally accessible quantity, such as the temperature
or luminosity of X-ray emitting intracluster gas, from which the 
mass can be determined via some physical model.
Relations between cluster mass and such a proxy typically require
the gas to be in virial equilibrium; however, many processes
are known that can disrupt the gas,
including cluster mergers
(\citealt{RS01}, hereafter RS; \citealt{RT02}; \citealt{PFB+06})
and AGN jet-blown radio bubbles \citep{MNW+05}.

There are many ongoing and planned surveys of clusters using
the Sunyaev-Zel'dovich (SZ) effect \citep{SZ72,Bir99},
which has the advantage of being effectively redshift-independent.
The SZ effect is proportional to the integral of the electron pressure along
the line of sight and can be characterized by the 
Comptonization parameter
  \begin{equation}
    y \equiv \frac{\sigma_T k_B}{m_e c^2} \int n_e T_e dl \propto \int P_e dl
    \, ,
    \label{eq:y}
  \end{equation}
where $n_e$ is the electron number density, $T_e$ is the electron temperature,
$P_e$ is the electron pressure,
and $l$ is the distance along the line of sight.
The actual SZ flux, measured as a decrement or increment in the
Cosmic Microwave Background (CMB),
depends on frequency and is subject to
relativistic effects for high temperature plasmas
\citep[for a review see][]{Rep95}.
Because we do not want to tie our results to any particular observational
project, we use the frequency-independent Comptonization parameter in 
the following study, as has been standard in the literature.
Also, we ignore any relativistic corrections as they are only relevant
for the most massive clusters and because they modify $y$ in a complicated
way that depends on frequency.

In general, SZ observations will give an image of the SZ effect or $y$ across
the cluster.
While specific values of $y$, for example the central or maximum value
for a cluster (hereafter 
$y_{\rm max}$), are not expected to be a particularly 
robust proxy for the mass, the integrated Comptonization parameter $Y$
displays a tighter correlation with mass \citep{RS06}.
This is defined as
  \begin{equation} \label{eq:Y}
    Y = \int y dA = d_{\rm A}^2 \int y d\Omega
    \, ,
  \end{equation}
where $A$ is the projected surface area of the cluster on the sky,
$\Omega$ is the solid angle, $d_{\rm A}$ is the angular diameter distance to the
cluster, and the integral is over the entire cluster on the sky.
Because the integrated Comptonization parameter is a global
quantity, proportional to $\int P_e dV$ or the thermal energy content of the
electrons,
it should be less sensitive
to non-equilibrium processes, which tend to be more localized in 
cluster cores.
The usefulness of SZ surveys to constrain cosmological parameters
has already been discussed extensively \citep[e.g.,][]{CHR02, HMH01, HHM01}.

As with X-ray proxies for mass, the regularity of an SZ--$M$ correlation
relies on the fact that many clusters are energetically close to
equilibrium.
However, dynamically unrelaxed clusters should add scatter to this 
correlation.
One mechanism known to disrupt the gas is cluster mergers,
a direct consequence of hierarchical structure formation.
How mergers affect the state of the gas will depend on the details 
of the individual mergers and their frequency, both of which
depend on the cosmological model.
To assess the utility of a mass proxy, such as the SZ effect, 
we need to quantify how mergers will affect the observed MF
and consequently the estimation of cosmological parameters.

Current cosmological simulations, which accurately trace the collapse
of structure and thus the merger history of clusters, cannot yet build
the large samples of clusters at sufficient numerical resolution to constrain
fundamental parameters and assess any potential bias due to mergers -- 
though this approach is becoming viable \citep[e.g.,][]{Hal+07}.
Typically, N-body cosmological simulations of dark matter are
re-simulated to include various types of gas processes such as ``preheating,"
radiative cooling, and AGN feedback, from which the scatter to an
observed SZ--$M$ correlation can be estimated.
Depending on the resolution of the re-simulated hydrodynamic grid,
these studies produce samples of $\sim10$ \citep{Nag06, BTD+07}
to $\sim100$ \citep{MHB+05, DKL+04} clusters.
Based on similar samples of simulations,
\citet{KVN06} have defined an SZ-like X-ray observable, $Y_X$,
which they have shown to be robust to nonequilibrium gas physics
with cosmological simulations.
Though suited to understanding the physical processes that add
statistical scatter to SZ--$M$ or similar relations, these samples are too
small to assess the effect of the scatter on the determination
of cosmological parameters, especially the effect of relatively 
rare, major merger events on the mass estimate of similarly 
rare massive clusters.
To include these rare events and focus expressly on the role of
mergers on SZ--$M$ relations and cosmological parameter estimates,
we take a semi-analytic approach that avoids simulating
every possible merger within a cosmological framework.

Specifically, we carefully examine the evolution of the SZ observables
$Y$ and $y_{\rm max}$ for a discrete set of detailed N-body/hydrodynamical
simulations of binary cluster mergers, generalize the results by
identifying and parameterizing the major transient features, or boosts,
and then apply these boosts to the merger histories of many clusters
generated semi-analytically via computationally cheaper merger trees.
We closely follow the methodology of
\citet[][hereafter RSR]{RSR02}, who similarly investigated the effect
of merger boosts on the X-ray observable mass proxies $L_X$ and $T_X$,
the X-ray luminosity and temperature respectively, and the bias such
boosts induce upon estimates of $\Omega_M$ and $\sigma_8$ from the 
inferred MFs.

To assess the impact of a particular world model or cosmology on our
results, we consider a ``flat" cosmology with a cosmological constant,
i.e. the $\Lambda$CDM concordance model, along 
with an ``open" and Einstein-de Sitter (``EdS") world model
for comparison;
the relevant parameters are summarized in Table~\ref{tab:omsig8_true}.
The dark energy equation of state and its evolution are only examined for the
flat universe.
The Hubble constant is parameterized as $100~h~{\rm km/s/Mpc}$ throughout.

In this paper, we assess the transient boosting of the SZ observables
$Y$ and $y_{\rm max}$ during cluster mergers, and the systematic influence
of mergers on cosmological parameter values derived from inferred
cluster MFs.
In \S~\ref{sec:sim} we describe the binary cluster merger simulations
from \citetalias{RS01} and the evolution of $Y$ and $y_{\rm max}$
during mergers.
In \S~\ref{sec:mergertrees} we discuss the generation of cluster
merger histories from merger trees created via the extended Press-Schechter
formalism \citep{PS74, LC93}, fit analytic functions that describe 
the transient behavior of merger boosts in the simulations, and
generalize these functions to the entire family of possible mergers.
In \S~\ref{sec:yM} the effect of boosts on the SZ--$M$
relations are analyzed, in \S~\ref{sec:yfs} the distribution function
proxies for the MF and the effect of boosts on them are described, and in 
\S~\ref{sec:fityfs} the distribution functions are used to assess
the impact of mergers on the cosmological parameters $\Omega_M$,
$\sigma_8$, and $w$.
Our results are discussed and summarized in \S~\ref{sec:concl}.

\section{Merger Simulations} \label{sec:sim}

To infer the effect of mergers on the SZ properties of clusters,
detailed N-body/hydrodynamical simulations of every
conceivable combination of cluster mass and impact
parameter would be ideal.
A realistic alternative is to use a small but representative set of
simulated mergers \citepalias{RS01} and interpolate or extrapolate 
from them the expected behavior of SZ observables for any set of 
merger parameters.

A detailed description of the simulations can be found in
\citetalias{RS01}.  
Eight simulated binary cluster mergers were
available with 3 mass ratios $M_{>}/M_{<}=$ 1, 3, and 6.5
each for 3 impact parameters $b=(0,2,5)r_{s}$ except the
$M_{>}/M_{<}=6.5$, $b=2r_{s}$ case.  
Here $r_{s}$ is the
scale radius in the NFW profile for the more massive
cluster \citep{NFW97}. 
In all simulations, the less massive cluster's
mass was fixed at $M_{<}=2 \times 10^{14} M_\odot$.  
Note
that the $M_{>}/M_{<}=6.5$ simulation runs are not specifically
mentioned in \citetalias{RS01}, although they were generated by the
same means as the other simulations.

\subsection{Equilibrium $Y$--$M$ and $y_{\rm max}$--$M$ Relations}
\label{sec:ynb} 

To compare the SZ properties of merging clusters with those
of similar clusters that are not undergoing mergers,
we need an equilibrium SZ--$M$ relation.
The theoretical models of clusters used in \citetalias{RS01}
are designed to represent observed,
non-cooling flow clusters and to have X-ray temperatures typical
of present day ``rich'' clusters.
These initial conditions therefore include ``preheating'' and
radiative cooling, though radiative cooling is ignored as a dynamic process
as the cooling timescale is designed to exceed a Hubble time.
Though cooling is absent in the \citetalias{RS01} mergers, our
results for $y_{\rm max}$ generally agree with a similar set
of cluster simulations \citep{PBM+07} that do include radiative cooling.
In any case, we are interested in the {\it change} of $Y$ 
or $y_{\rm max}$ due to mergers and not in precisely characterizing
the equilibrium state of clusters.
To accurately assess the relative effect of mergers on the SZ effect,
we take the initial clusters in \citetalias{RS01} 
as our equilibrium clusters, which should correspond well to actual clusters
since they were built to resemble observed, relaxed clusters. 

Self-similar scaling
relations derived from virial arguments \citep{DKL+04, CK05}
give $Y \propto M^{5/3} f_{g}$, where $M$ is the virial mass
and $f_{g}$ is the gas mass fraction.  
For masses $M \gtrsim 10^{14}
M_{\odot}$, $f_{g} \propto M^{1/3}$ though $f_{g}$ steepens at
smaller masses.
This general trend of increasing $f_g$ with mass has been observed 
for relaxed, nearby clusters \citep{VKF+06}.
Assuming all clusters have similar density profiles,
we find that
$y_{\rm max} \propto M f_{g}$.  
For the initial clusters in the simulations,
we calculate exact solutions for $Y$ and $y_{\rm max}$:
  \begin{equation} \label{eq:Yintdef}
    Y = 0.210 \left( \frac{M}{10^{15} M_{\odot}} \right)^{2}
    \left( \frac{r_{s}}{{\rm kpc}} \right)^{-1}
    f_{g} \, h^{-2}~\mathrm{Mpc}^{2}
    \, ,
  \end{equation}
  \begin{equation} \label{eq:ymaxdef}
    y_{\rm max} = 8.84 \times 10^{3}
    \left( \frac{M}{10^{15} M_{\odot}} \right)^{2}
    \left( \frac{r_{s}}{{\rm kpc}} \right)^{-3}
    f_{g}
    \, .
  \end{equation}
Here, $r_s$ and $f_{g}$ are found numerically 
\citepalias[equations~(20)-(23),][]{RS01}.  
Over the range of cluster masses we consider, $Y$ and $y_{\rm max}$ scale
approximately as $Y \propto M^{2}$ and $y_{\rm max} \propto M^{1.3}$.

In practice, we fit the numerical solutions for $Y(M)$ and
$y_{\rm max}(M)$ each to a power law times a 13 degree polynomial.
The high order of the polynomial is required primarily because we
need the derivatives of the function to compute the $Y$ and $y_{\rm max}$
distribution functions (YF and yF respectively).  
The fractional error in the derivatives
of the fits is $\lesssim 1\%$ for both $Y(M)$ and $y_{\rm max}(M)$, and better
than that for the fits themselves.

\subsection{Merger Boosts to $Y$ and $y_{\rm max}$} \label{sec:yb}

\subsubsection{Generating $Y$ and $y_{\rm max}$ from the
Simulations} \label{sec:yt}

For each simulation in \citetalias{RS01}, the
behavior of the X-ray temperature and luminosity was
calculated
\citepalias[see][Figures~5 \& 8]{RS01} as a function of time.
We would like similar curves for $Y$ and $y_{\rm max}$;
however, these quantities were not calculated during the
simulations, so we need to evaluate them from saved 3D
``snapshots" of the simulation grid in order to recreate the
evolution with time.  
For most of the runs, 40 to 60
snapshots were saved fairly regularly over the 14 Gyr
the mergers were followed.  
From the gas pressure distribution,
the Comptonization parameters can be calculated individually for each
snapshot and combined to trace the evolution of $Y$ and $y_{\rm max}$
during the merger.

\begin{figure}
\plottwo{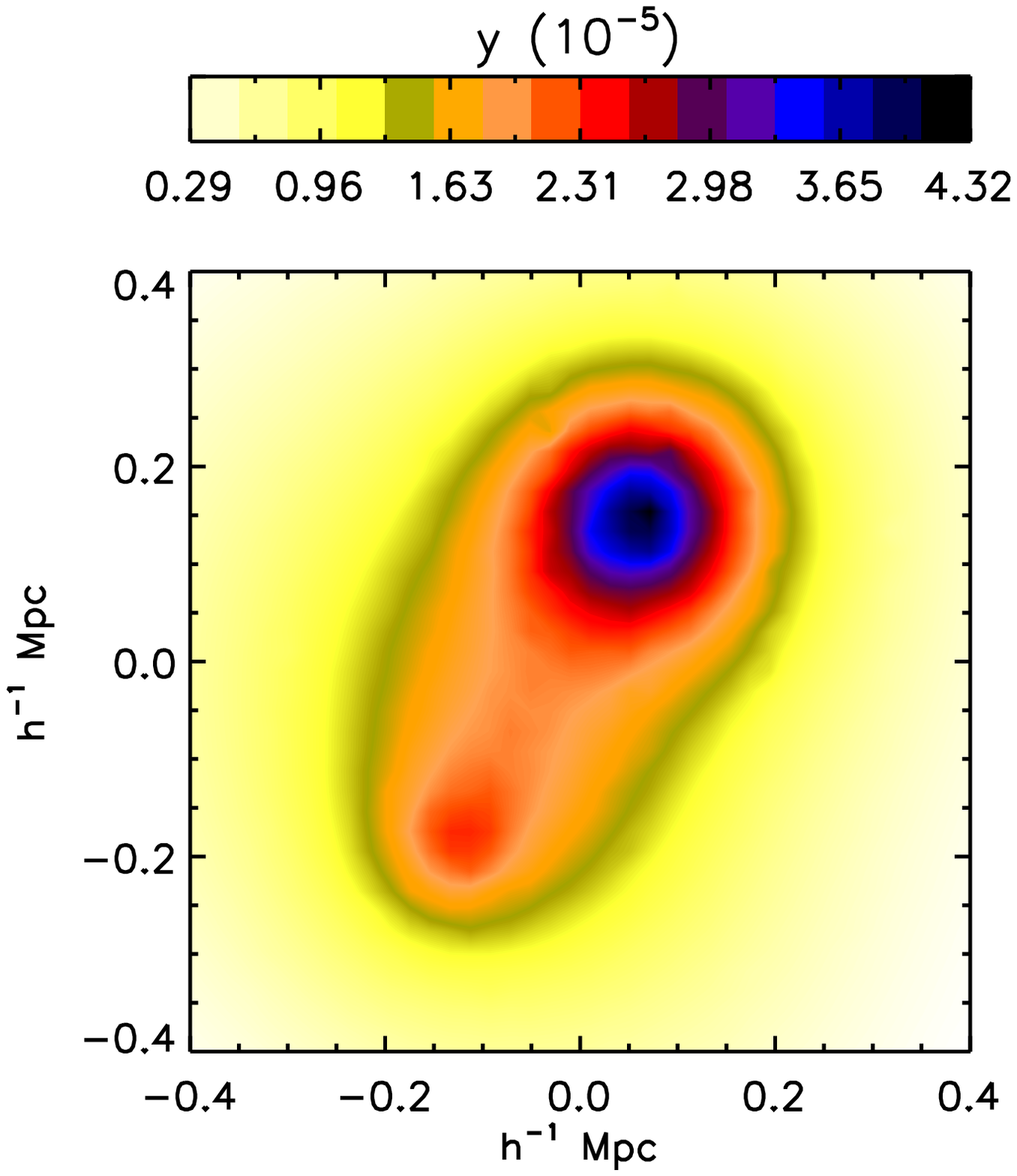}{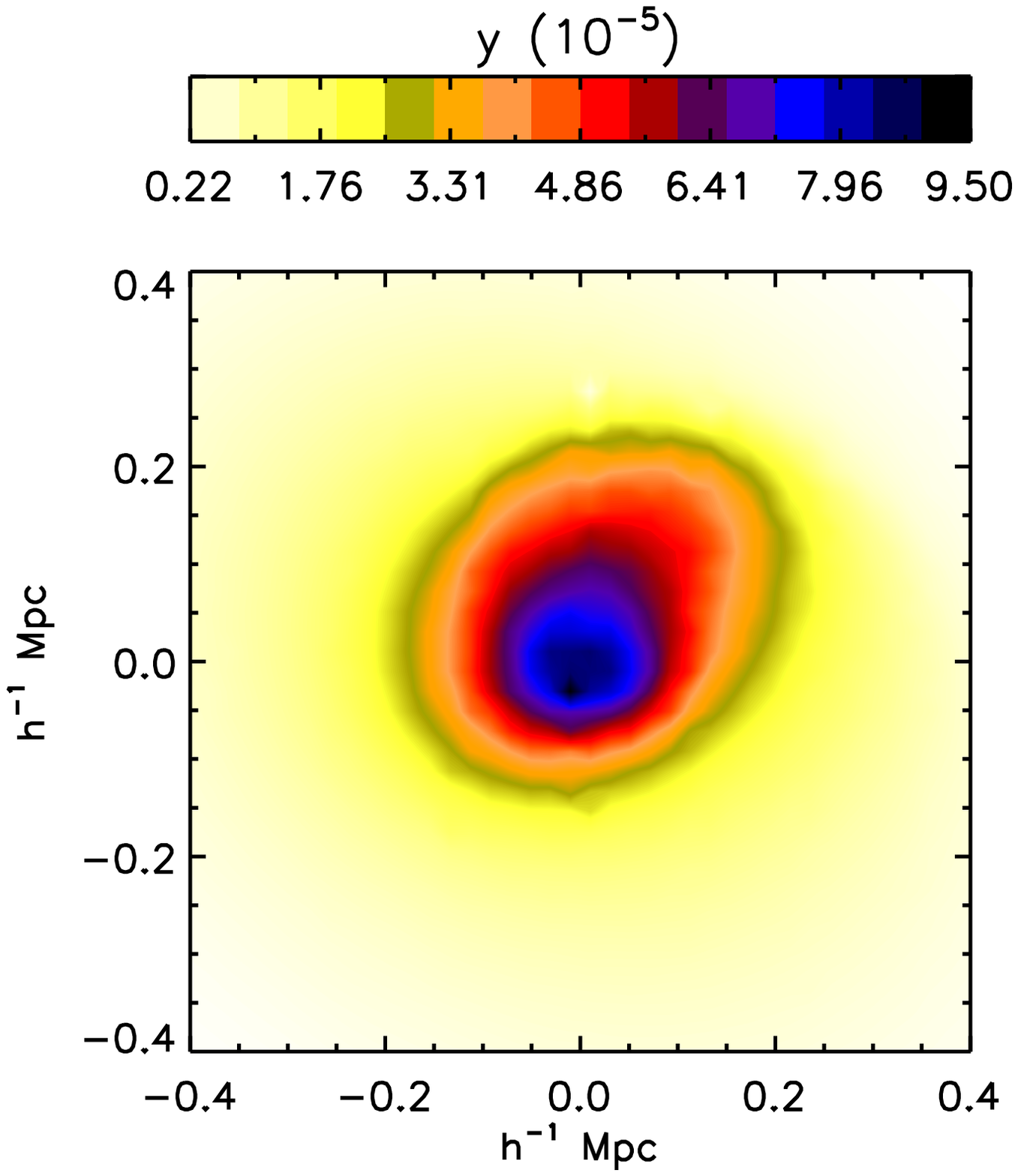}
\caption{Images of the SZ parameter $y$ from 3D snapshots of the
1:3 mass ratio, 2 $r_s$ impact parameter merger simulation.
Here $r_s$ is the NFW scale radius of the more massive cluster.
Both images are viewed from a line-of-sight which is rotated $45^\circ$
from the merger axis and $45^\circ$ azimuthally from the merger plane.
{\it Left:} 386 Myr before first core crossing.
{\it Right:} 114 Myr after first core crossing.
\label{fig:yim}}
\end{figure}

Simulated SZ images for any orientation can be generated for
each snapshot.
As an example, Figure~\ref{fig:yim} shows $100\times100$ pixel images 
from 2 snapshots of the $M_>/M_<=3$, $b=2r_s$ merger.
For both of these images, our line-of-sight is oriented at $45^\circ$
to the merger axis and rotated
$45^\circ$ azimuthally from the merger plane.
In this particular example, the clusters are seen just before and
just after the first core crossing, which generally corresponds 
to the maximum transient enhancement of both $Y$ and $y_{\rm max}$.
Note that while the images look qualitatively similar, the scale of
the image after core passage is twice that of the pre-core passage
snapshot, suggesting that both $Y$ and $y_{\rm max}$ should get
``boosted" during a merger event.

To compute 
$Y = \int y dA = \frac{\sigma_T}{m_e c^2} \int P_e dV$, 
we simply add up
the pressure in each computational cell weighted by the cell volume so that
$Y = \frac{\sigma_T}{m_e c^2} \displaystyle \sum_i P_{e,i} \Delta V_i$,
where the sum is over all the cells in the 3D grid.  
We do not restrict the
integration to the virial radius $r_{200}$ or $r_{500}$
as in other cases where $Y$
has been modeled
\citep{DKL+04, MHB+05}
for several reasons.  
First, the
initial conditions for the simulated clusters cut off the
pressure and density profile at the virial radius,
so these definitions are at least initially equivalent. 
Also, during the merger there is no such well-defined radius as
the gas is interacting violently.
However, nearly all of the contribution to $Y$ comes from gas inside
the virial radius: 99.5\% initially and 95\% after the clusters have
merged and equilibrated.

For each snapshot of each merger simulation, values of $y_{\rm max}$ are
computed
for 339 orientations of the merger relative to our line of sight.
Because the effects of the merger on the value of $y_{\rm max}$ tend to 
vary the most near the merger axis, we more finely sampled the viewing
angles in this direction.
The orientations sampled with respect to the merger axis are uniformly
spaced in $\sin\theta$, where $\theta$ is the polar angle,
such that $\Delta\sin\theta=1/15$.
The sampling of the azimuthal angle $\phi$ is varied, to ensure relatively
even spacing, as $\Delta\phi=8^\circ/\sin\theta$.
To determine $y_{\rm max}$ for
each merger, snapshot, and orientation, values of $y$ were computed
by integrating along 16 lines of sight (equation~(\ref{eq:y})), on a
4x4 grid, to form an SZ image of the cluster as seen
from that orientation.
The grid was then recentered on the maximum value of $y$ and reduced
in scale by a factor of 3.5, and $y$ was calculated again.
This procedure was repeated until the maximum value on the grid 
varied by less than 0.1\% compared to the value from the previous
iteration, and this
$y$ is adopted as $y_{\rm max}$. 

\subsubsection{Correcting for Mass Loss Outside the Grid} \label{sec:massloss}

During each merger, some gas is flung out to large radii and lost
from the simulation due to the finite size of the computational grid and
outflow boundary conditions at the grid edge.  
Of course, once
the gas is outside the simulation grid, it is permanently lost.
Noticeable amounts of gas do not leave the grid until after
the first core passage.
Since we
are mainly interested in the times when the merger boost is
large, which occurs near the peak associated with first core passage,
our results are not particularly affected by the lost gas.  
At late times,
however, after the clusters have merged, $Y$ remains below
the expected value for a cluster with mass
$M_{\rm total} = M_< + M_>$.
Since at these cluster masses
$Y \propto M^2$ and $y_{\rm max} \propto M^{1.3}$, we correct
for the lost gas by taking
  \begin{equation} \label{eq:YM}
    Y =
    \left[ \frac{M_{\rm gas}(t=0)}{M_{\rm gas}(t)} \right]^2
    Y_{\rm calc} \, ,
  \end{equation}
and
  \begin{equation} \label{eq:yM}
    y_{\rm max} =
    \left[ \frac{M_{\rm gas}(t=0)}{M_{\rm gas}(t)} \right]^{1.3}
    y_{\rm calc}
    \, .
  \end{equation}
Here, $Y_{\rm calc}$ and $y_{\rm calc}$ are the integrated and 
maximum SZ parameters
calculated by integration over the grid prior to this correction.
The correction is small;
over the duration of the first peak in $Y$, which
is much longer than the peak in $y_{\rm max}$, less than
5\% of the gas has been lost from the grid.

In fact, some of the lost gas exits the grid near to or above escape velocity,
assuming a collisionless ballistic trajectory, so correcting for its
loss may seem inappropriate.
The majority of the gas, except during the short period after the 
first core passage, effectively leaks out of the grid due to a lack of 
pressure support at the simulation boundary.
This artificially lower pressure propagates inward, requiring the correction
we apply; otherwise, the boost will be slightly underestimated.
After clusters have formed in cosmological numerical simulations, the
gas fraction at the virial radius is generally 10\% below the cosmic
baryon fraction \citep{CEF+07, ENF98}, 
perhaps indicating that up to $\sim$5\% of the gas
has been ejected, given that 5\% of the baryons are in stars.
The simulations of \citetalias{RS01} we utilize cannot accurately follow
the merger to its true final state and so we cannot address the question
of true gas ejection from clusters after merger events.
However, the initial simulated clusters are constructed to match
observed clusters with realistic gas fractions, 
so if gas is in fact lost, that effect is intrinsically
included by \citetalias{RS01} and the resultant boosts in $Y$ and $y_{\rm max}$.
Our conclusions are not drawn from any late time evolution in the
simulations, nor do we investigate the true post-merger state of
clusters.

Additionally, we correct $Y$ for the slight evolution at large radii
in the relaxed, pre-merger profiles of the simulated clusters.
Because the integrated Comptonization parameter
is inversely proportional to a low power of cluster radius due to 
$Y \propto n_e$, the outer parts of a cluster contribute
significantly to its overall value, as compared to $L_X$, which is
proportional to $n_e^2$.
The lower pressure in the
outer regions can affect $Y$ because there is more volume at large
radii, even though $y_{\rm max}$ remains unaffected.
We observe a slight drop in the
pressure profile outside the central core over time before
the individual clusters begin to interact, which is likely
due to the artificial truncation of gas at the virial
radius -- gas at this boundary is not in hydrostatic
equilibrium in the simulations and will flow outward, and the loss of pressure
support will travel inward, readjusting the profile as
the system tries to establish hydrostatic balance.
While for the least massive cluster this effect is hardly noticeable,
the magnitude of the effect increases with total cluster mass.  
Fortunately, the effect on $Y(t)$ appears
to be linear in time, so we correct the time evolution of $Y$
such that $Y$ is forced to be constant before the
clusters begin to interact, normalized to $Y(t=0)$.

\subsection{Evolution of $y_{\rm max}$ and $Y$ During
Mergers} \label{sec:discy}

In Figure~\ref{fig:yt},
$Y$ and $y_{\rm max}$ are shown as a
function of time for the merger simulations including the corrections
described in \S~\ref{sec:massloss}.
For $y_{\rm max}$, the plot is shown for a viewing orientation at
90$^\circ$ to the merger axis and in the merger plane.
The maximum boost for the head-on collision in $y_{\rm max}$ is nearly 
a factor of 10,
while the boost in $Y$ is always less than a factor of 2,
though the duration of the boost in $Y$ is much longer
than that for $y_{\rm max}$.
\citet{MHB+05} report a maximum boost factor in $y_{\rm max}$ of 20
in cosmological simulations re-simulated to include gas hydrodynamics,
twice the amount of boosting we find, 
though their result could be due to an artificially
high central temperature in their pre-merger clusters \citep{LNN+02}.
However, it is more likely the enhanced boost is due to the
natural inclusion of multiple mergers and constant accretion along
filaments, which are not included in binary merger simulations.
For example, a triple merger between 2 equal mass clusters and a third
subcluster with a tenth of one of their masses should yield a boost factor of
20, extrapolating our results to such a case.
Additional pressure due to bulk motions within the pre-merger clusters,
producing stronger shocks, may also lead to a larger boost.
Globally, the temperature profile of the initial clusters in \citetalias{RS01}
agrees well with those clusters assembled in cosmological simulations
\citep{LNN+02}, so the precise origin of the discrepency is unclear.
However, our boost factors are confirmed
in a recent set of binary cluster mergers \citep{PFB+06}, in which
\citet{PBM+07} find $y_{\rm max}$ to be boosted by a factor of 
$\sim10$ (see their Figure~7).

Essentially, $y_{\rm max}$ traces the densest parts of clusters, which are
the cores.
These remain reasonably intact until near the time of first core crossing,
which makes the peak in $y_{\rm max}$ relatively narrow.
On the other hand, $Y$ involves a sum of all the gas, so it begins to get
boosted as soon as gas at large radii starts to interact,
long before the cores approach, and the boost lasts longer, as gas
in the outer regions needs more time to re-equilibrate.
The time evolution of $y_{\rm max}$ is qualitatively similar to 
that found by
\citetalias{RS01}
for the X-ray temperature and 
luminosity, quantities that are also dominated by the
cores of clusters
due to the fact that the X-ray emissivity depends on the square of the
density.

The plots of the evolution of $Y$ already indicate that this
parameter will not be strongly affected by mergers.
First, the boosts in $Y$ are smaller than in $y_{\rm max}$.
Second, the boosts are not large compared to the equilibrium effect of
increasing the mass.
Assuming $Y\propto M^2$, the boost factor
$B$ needed to exceed the final equilibrium value of $Y_{final}$
is $B > \frac{Y_{final}}{Y_1 + Y_2} = \frac{(M_1+M_2)^2}{M_1^2 + M_2^2}$.
For equal mass mergers, this condition gives $B > 2$,
and from Figure~\ref{fig:yt} it is clear the boost factor is always $<2$.
If $Y$ is used as a proxy to determine the mass of a cluster, the
resulting value during the merger will nearly always lie between the 
individual initial masses
of the subclusters and the final total mass.
In a certain sense, this only affects the definition of when the cluster has
merged, and the applicable mass, and does not represent a real bias.
We find that other mass ratios 
can boost $Y$ beyond the final equilibrium value, but only by
factors slightly larger than unity.

\section{Merger Trees} \label{sec:mergertrees}

Structure formation and evolution are most easily traced through
the mass function (MF) of dark matter halos, $n(M, z)$, where
$n(M, z)dM$ gives the number of halos per unit comoving volume
with masses in the range $M \to M+dM$.  
Currently, the MF for
a given cosmology at a given redshift can be found most
accurately from numerical N-body simulations \citep{Spr+05}.
While accessing the results of these simulations has become
more feasible \citep[e.g.,][]{LS06}, a semi-analytic approach
to obtaining the MF proves more practical, especially since
we are concerned with the relative effect of
merger boosts on the underlying MF and not the precise nature
of the MF itself.  
We follow the PS formalism, which agrees
with the MF found in numerical simulations, especially
at higher masses \citep{BN98};
specifically, we use extended Press-Schechter theory
as developed in \citet{BCE+91} and \citet{LC93} and applied in
\citetalias{RSR02}. 
Though the PS formalism fails to reproduce the MF found in numerical
simulations at very high redshifts and low cluster masses 
\citep[see, e.g.,][]{ST99, LHH+07}, it is more than sufficient over the
redshifts ($z=0\to2$) and masses ($M=10^{14}\to10^{16}M_\odot$)
of interest here.

\citet{PS74} give the MF at some redshift $z$ as
  \begin{equation} \label{eq:nps}
    n_{PS}(M,z)dM = \sqrt{\frac{2}{\pi}}
    \frac{\overline \rho}{M} \frac{\delta_c(z)}
    {\sigma^2 (M)} \Bigg| \frac{d\sigma(M)}{dM} \Bigg|
    \mathrm{exp} \Biggl[ - \frac{\delta_c^2(z)}
    {2\sigma^2(M)} \Biggr] dM
  \end{equation} 
where
$\sigma(M)$ is the current rms density fluctuation within
a sphere of mean mass $M$, and $\delta_c(z)$ is the critical
linear overdensity required for a region to collapse at redshift $z$.
The derivation of this expression assumes that halos grow
from Gaussian density fluctuations that have
larger amplitudes on smaller scales.  
Structure then forms
hierarchically, with small halos collapsing first and
merging to form larger halos.  
In this scenario, the highest
mass halos, observed as clusters of galaxies, form
most recently and should be most affected by merger
processes at the present day.

From this extended PS formalism, we follow the
procedure outlined in \S~3 of \citetalias{RSR02}, in which
a ``merger tree" is generated for a present day cluster.
The merger tree traces the merger and accretion history
of a cluster of mass $M$ back in time.  
For each time step,
a progenitor cluster of mass $M_{p1}$ is chosen from a probability
distribution \citep[][equation~(2.25)]{LC93}, and since we
only consider binary mergers, the mass of the other
progenitor cluster is given by $M_{p2}=M-M_{p1}$.
We will use the notation $M_>$ and $M_<$ for the larger and smaller masses of
the subclusters in each binary merger.  
\citetalias{RSR02}
in \S~3.1 discusses the disadvantages of dealing solely with binary
mergers; however, our set of simulated mergers does not
address more complex mergers, so we have no good way to
derive a boost for them.  
Also, boosts are most dramatic for
near equal mass mergers, and in such cases additional merger
participants will likely be much less massive and have a negligible
effect on the resultant boosts.
However, one result of ignoring multiple mergers is that the merger
tree-derived MF tends to overestimate the analytic PS MF for $z>0$.
The progenitor cluster with mass $M_{p2}$ is not taken from the PS
distribution and is generally overestimated, so that the high mass end
of the MF is overestimated at the expense of the very low mass end.
Since we concern ourselves with the highest mass clusters, our resulting
MFs will lie slightly above the analytic prediction, as illustrated in
Figure~\ref{fig:bnbyF}, which we must
take into account when fitting MFs in \S~\ref{sec:fityfs}.

A large number of merger trees was created with a broad span of initial
cluster masses, 
and the distribution of the initial masses was weighted so as to
give the present day mass function.
From ensembles of merger trees for the cosmologies of
interest, we can find the MF at any redshift, and at any redshift we
have each cluster's merger history, which can be used to determine
the merger boost in some observable --- in our case $Y$ and
$y_{\rm max}$.

Merger trees are a simple and computationally cheap way to
simulate structure formation for a particular world
model.  
But they are limited in that they only specify
progenitor cluster masses and discrete time intervals
during which the mergers occur.  
All the dynamics
and other details of a merger, however, are absent from
EPS-derived merger trees.  
The information needed
to connect the trees to our merger simulations is the masses of the
clusters, the impact parameter $b$ of the encounter, and the
time of first core passage, which we designate as the time
of the merger, $t_{\rm merge}$, in the merger trees.
While the masses are provided by the merger trees,
an appropriate $b$ must
be selected for each merger in the trees.  
We follow the method in \S~6 of \citetalias{RSR02}, where
a value for the spin parameter is chosen from a
Maxwell-Boltzmann-like distribution, which represents the
observed distribution from numerical simulations \citep{BDK01},
allowing $b$ to be derived from the chosen spin parameter
\citep{Sar02}.  
To determine the precise value of
$t_{\rm merge}$, we simply select a
random time within the small discrete time step used in the merger trees,
and take that time to be the instant of first core passage,
since the merger could have occurred at any point within that time.

\subsection{Merger Boost Histograms} \label{sec:histograms}

As discussed in \citetalias{RSR02},
the effect of a merger boost on a cluster whose history is characterized by
a merger tree can be determined from a histogram which gives the
magnitude of the boost as a function of time.
Since the merger trees give a statistical description of the history of
cluster mergers, it is sufficient to determine the distribution histogram
of boosts versus the observed time $t_{\rm obs}$.
The form of the histogram reduces the details
contained in the curves in Figure~\ref{fig:yt} to a simpler,
one-to-one function
that can be fit by the merger
parameters $M_{<}$, $M_{>}$, and $b$.  
In the fits, we scale
the impact parameter $b$ by the core radii of the two merging clusters,
$b' = b/(r_{c<}+r_{c>})$ and the time by the ratio of the virial
radius of the more massive cluster to the gas sound speed, $t_{\rm sc}$.
A more detailed explanation of these scalings
is given in \S~5.3 of \citetalias{RSR02}.

\subsubsection{Fitting $Y$ Histograms from Simulations}
\label{sec:Yinthist}

The left panel of Figure~\ref{fig:histy} shows the cumulative time spent
by the system above any given value of $Y$
for the $M_>/M_<=1$ merger simulations.
We use cubic spline interpolation of the boost curves 
in the left panel of Figure~\ref{fig:yt} to produce a smoothly
varying histogram.  
In \citetalias{RSR02}, the $T_X$ and $L_X$ boost histograms were well-fit
by hyperbolas parameterized by the equations given in their
Appendix B.
We find hyperbolas also well-describe the $Y$
histograms, and we use the same parameterization as
\citetalias{RSR02}
with only a minor change given in Appendix~\ref{sec:appendix}.

\subsubsection{Fitting $y_{\rm max}$ Histograms from Simulations}
\label{sec:ymaxhist}

The procedure for $y_{\rm max}$ is slightly more complicated
due to the orientation-dependence of the central Comptonization parameter.
The evolution of the
maximum value of $y$ as a function of viewing angle varies more 
dramatically near the merger axis than perpendicular to it.  
As noted in \S~\ref{sec:yt}, for each
merger simulation, the evolution of $y_{\rm max}$ with time was 
calculated for 339 orientations, sampling more finely
around the merger axis.
Because $y_{\rm max}$ really
traces the cluster cores, the peaks in the curves
are larger and have shorter durations.
As a result, we found that simple interpolation did not sufficiently 
sample the peaks,
so we use a superposition of Gaussians to fit the shape of the boost 
as a function of time.

The merger trees contain no information about the orientation 
of the cluster mergers, and
we assume an isotropic distribution relative to our line-of-sight.
For our grid of 339 viewing angles, the probability of any one orientation is 
determined by the solid angle of that grid cell.
We weight each orientation by this solid angle divided by $4 \pi$.
All 339 $y_{\rm max}(t)$ curves, weighted by their probability
of being observed, are used to construct a histogram like those described
in \S~\ref{sec:Yinthist}.
The histograms for the $M_>/M_<=1$ runs are shown in the 
right panel of Figure~\ref{fig:histy}.
Since these histograms include the distribution of merger boosts for all
orientations of the line of sight, the boosts need to be normalized to
the pre-boost value of $y_{\rm max}$ for some fixed orientation.
The boosts in the right panel of Figure~\ref{fig:histy} were taken relative
to the pre-boost $y_{\rm max}$ as observed 90$^\circ$ to the merger axis and
in the merger plane.
Note that this is the same orientation assumed in the right panels of
Figure~\ref{fig:yt}.

Because the $y_{\rm max}$ histograms include the results from many 
different orientations,
the high boost ends of the histograms decline more slowly with time
than for $Y$ or $L_X$ or $T_X$.
A different function is thus used to fit these
histograms (see Appendix~\ref{sec:appendix}).  

\subsection{Generalizing Merger Boosts for Arbitrary Mass
Ratio and Impact Parameter} \label{sec:fitallhists}

As in \citetalias{RSR02},
the parameters of the fits to the boost histograms were fit to
simple functions of the masses and impact parameter in the merger.
The forms of these functions were chosen so as to have the correct
asymptotic forms (e.g., in the limit of large $M_>/M_<$).
The free parameters of these functions were chosen to
best fit the histograms from all 8 simulation runs. 
The values of these parameters 
are given in Table~\ref{tab:appendix} (below in the Appendix) for the $Y$ and 
$y_{\rm max}$ histograms.  

The maximum fractional error in the fits to the boost simulation data 
for $Y$ is $<3\%$ except for the 2 runs with $M_{>}/M_{<}=6.5$.  
Here, the evolution of the pressure
distribution in the more massive cluster before collision
dominates the time evolution of $Y$ (see \S~\ref{sec:massloss}).  
The fits overestimate the boosts for the $M_{>}/M_{<}=6.5$ simulations;
however, the boosts themselves are small in this case, and the errors are
still $<10\%$.  

For the $y_{\rm max}$ fits, the average fractional error is
typically 4\%, and the maximum error is $<10\%$.
We found that the time sampling for the $M_{>}/M_{<}=3, b=2r_s$ simulation
run was too sparse around the boost to strongly constrain the shape 
of the $y_{\rm max}$ histogram, so we did not use this run in our fits.

\subsection{Adding Boosts to Merger Trees} \label{sec:applymt}

With the fitted forms for the histograms for the strength of a boost
versus time as a function of
the masses of the merging subclusters $M_<$ and $M_>$ and the
impact parameter $b$,
the boosted values of $Y$ or $y_{\rm max}$
can easily be found for clusters from their past merger histories given by
the merger trees.
For any redshift or observed time, $t_{\rm obs}$, we search back through
a cluster's merger tree and for every merger event, we find the 
boosted value of $Y$ or $y_{\rm max}$ for that merger.
If the boosted $Y$ or $y_{\rm max}$ exceeds the value given by our
equilibrium equations~(\ref{eq:Yintdef}) and (\ref{eq:ymaxdef}) for 
the mass of the cluster at $t_{\rm obs}$, then we assign the boosted
value to that cluster's observed $Y$ or $y_{\rm max}$; otherwise
it acquires its equilibrium value.
Boosted values less than those given by the equilibrium equations are
not allowed because
the analytic fits from which boost factors are derived poorly describe
the histograms, such as those shown in Figure~\ref{fig:histy}, for
negative and small positive boosts.
While the discrepency between the simulation-based histograms and the analytic
fits for small boosts leads to an underestimate of the number of these
clusters, we are primarily concerned with the more dramatic
effects caused by large boosts, which are well-described by the fits.

\section{SZ versus Mass Correlation} \label{sec:yM}

Once clusters observed at some redshift are assigned values of 
$Y$ and $y_{\rm max}$ based on each cluster's merger history,
we can evaluate the robustness of the $Y-M$ and $y_{\rm max}$--$M$
relations.
The top panels of Figures~\ref{fig:yiscat} and \ref{fig:ymscat}
show $Y$ and $y_{\rm max}$ versus mass
for clusters in our merger trees at $z=0$ and $z=1$.
Most clusters have nearly unboosted values of $Y$ and $y_{\rm max}$,
while the number of clusters that deviate from either SZ--$M$ relation drops
roughly exponentially with the strength of the boost.
We find that $\sim 15\%$ of clusters are boosted in $y_{\rm max}$
by $\ga 15\%$ and in $Y$ by $\ga 0.1\%$.
Note that the scatter in Figures~\ref{fig:yiscat} and \ref{fig:ymscat}
is due entirely to merger boosts and does not include observational error or
scatter related to other physics.

As expected, many clusters are found to have significantly boosted values of
$y_{\rm max}$, which overestimate the actual masses.
However, there are almost no ``boosts" to $Y$ in Figure~\ref{fig:yiscat}.
Instead, we see clusters scattered {\it below} the $Y$--$M$ relation, as
is also seen, though to a lesser extent, in Figure~\ref{fig:ymscat}.
Clusters that fall below the $Y$--$M$ relation were ``observed" after
a merger (after the peak of the boost), but before virialization.
It should be noted that, according to \S~\ref{sec:applymt}, a
boost is only applied if it gives a $Y$ or $y_{\rm max}$
greater than its equilibrium value {\it before} the merger,
while in Figures~\ref{fig:yiscat} and \ref{fig:ymscat} the mass is
taken to be the final, or merged, mass of the clusters.
So, though clusters can never fall below their pre-merger equilibrium 
relation in our formulation, a boosted cluster may fall below its 
post-merger value.
The scatter below the $y_{\rm max}$--$M$ relation is less pronounced 
due to the shorter period when the SZ effect is below the eventual
equilibrium value (see Figure~\ref{fig:yt}).
This feature is a general characteristic of observing a recent post-merger
cluster and will be difficult to identify as such in an actual survey, 
and will likely affect the normalization of either SZ--$M$ relation.

In order to quantify the effect of mergers on the SZ versus mass
relations, we fit power-law functions of the form
\begin{equation} \label{eq:YvsM}
Y = A \times 10^{-5} \, h^{-2} \,
\left( \frac{M}{10^{15} \, M_\odot} \right)^\alpha \, {\rm Mpc}^{2} \, ,
\end{equation}
or
\begin{equation} \label{eq:yvsM}
y_{\rm max} = A \times 10^{-5} \,
\left( \frac{M}{10^{15} \, M_\odot} \right)^\alpha \, ,
\end{equation}
to all of the clusters with
$Y>10^{-5} h^{-2} {\rm Mpc}^2$ or $y_{\rm max}>10^{-5}$.
We estimate the scatter with respect to the best fit, 
$\sigma_{\rm fit}$, and also the scatter and offset with respect to the
actual equilibrium relations for the SZ effect
(equations~(\ref{eq:Yintdef}) and (\ref{eq:ymaxdef})), $\sigma_{\rm eq}$.
We define the scatter as
  \begin{equation} \label{eq:scatter}
    \sigma^2_{\rm fit} = \frac{\displaystyle \sum_i (y_i-y_{{\rm fit},i})^2
    /y_{{\rm fit},i}^2}{N-1} \, ,
  \end{equation}
and $\sigma_{\rm eq}$ is similarly defined, except $N-1$ is replaced by $N$.
The coefficients $A$ and $\alpha$ of the fits along with the scatter are
given in Table~\ref{tab:scatter}.
The subscripts ``b" and ``nb" refer to clusters including merger boosts, and
not including these boosts (where the SZ properties are given by
the equilibrium relations).
Note that we consider a logarithmic distribution as in \S~\ref{sec:yfs}
for the cluster masses, so that the fits in Table~\ref{tab:scatter},
as well as the points in Figures~\ref{fig:yiscat} and \ref{fig:ymscat},
do not reflect the actual MF of clusters.

Because the relative strength of boosts is mainly a function of mass ratio 
and is only weakly dependent
on the absolute masses of the merging clusters, clusters are boosted 
somewhat uniformly in $Y$ and
$y_{\rm max}$ across masses, which tends to change the 
normalization of the fit, $A$, and only to a lesser extent the slope, $\alpha$.
The inclusion of merger boosts, in the case of $y_{\rm max}$, 
could either flatten or steepen the slope.
The local mass function is flatter at the low-mass end, so
low-mass clusters experience more high-mass-ratio mergers
overall than high-mass clusters, thus flattening the
$y_{\rm max}$--$M$ relation.
However, when both minor and major mergers are considered, at any given
time the high-mass clusters are undergoing more merger events
(see the relative change in yF over time for low- and high-mass
clusters in Figure~\ref{fig:bnbyF}).
Thus, at any given time
a higher-mass cluster has a greater probability of finding itself
in the midst of a merger of some type.
If the mass function
is oversampled at the high-mass end, this effect tends to
steepen the $y_{\rm max}$--$M$ relation.
Because our cluster sample has a uniform
distribution in log mass, we oversample the high-mass end relative
to the low-mass end.
Consequently our $y_{\rm max}$--$M$
relation does not exhibit the flattening that we would expect
if our cluster sample had been drawn from the correct mass function.
As Figure~\ref{fig:ymscat} shows, both high- and low-mass clusters
exhibit the same number of large boosts, but the total number of
boosted clusters is greater at higher masses.  Most of the
high-mass clusters with boosts have small boost factors that are
difficult to see in the figure.

The boosted normalization for $Y$ is systematically lower than the
nonboosted $A$, but by $<1\%$.
In the case of $y_{\rm max}$, the normalization increases by $\sim 10\%$.
The offsets to $y_{\rm max}$ are due as much to clusters with
small boost factors as to the rarer cases with very large boosts.
Note that these clusters tend to be undergoing weaker mergers, which
may be hard to detect.
Thus, it may be difficult to expunge these clusters
from SZ surveys, and the systematic shift in the $y_{\rm max}$ versus
mass relation may bias cluster samples.
The merger-induced scatter to the fit, $\sigma_{\rm fit}$, is $\sim 2\%$
for the $Y$--$M$ relation and 25-30\% for the $y_{\rm max}$--$M$
relation, and is nearly independent of the cosmological world model and redshift.
The scatter relative to the equilibrium relations, $\sigma_{\rm eq}$,
increases with redshift since the merger rate is higher in the past,
whereas the addition of boosted clusters adjusts the normalization $A$
to minimize $\sigma_{\rm fit}$, so the scatter remains about constant between
redshifts.
Also, because there are fewer clusters that show boosts in $Y$, 
the scatter $\sigma_{\rm fit}$
is dominated by deviations of the equilibrium relation from a power-law form,
which explains why $\sigma_{\rm eq}$ tends to be slightly smaller than
$\sigma_{\rm fit}$ for $Y$.

\section{Distribution Functions of $Y$ and $y_{\rm max}$} \label{sec:yfs}

We computed the distribution functions for $Y$ and $y_{\rm max}$,
which we refer to as the YF and yF, respectively.
The distribution function YF is $n( Y, z )$, where
$n( Y, z )~d Y$ gives the number of clusters per unit comoving volume 
at redshift $z$ which have integrated SZ parameters in the range
$ Y \rightarrow Y + dY$.
The yF distribution function $n( y_{\rm max}, z )$ is defined in an 
equivalent manner.
To build the YF from a merger tree, we find all the clusters that
exist at the ``observed'' redshift and assign a value of
$Y$ according to \S~\ref{sec:fitallhists} for the non-boosted YF and
\S~\ref{sec:fitallhists} for the boosted YF.
A cluster with integrated Comptonization parameter $Y_i$
is then added to a pre-determined bin ${\rm YF}_j$ such that
$Y_j^{\rm bin} \leq Y_i < Y_{j+1}^{\rm bin}$
and appropriately weighted to convert the actual initial distribution
of $z=0$ cluster masses used in the merger trees, $dN/dM^0$, to the
Press-Schechter distribution $n_{PS}$:
  \begin{equation} \label{eq:yf_mt}
    {\rm YF}_j = {\rm YF}_j + \frac{n_{PS}(M^0, z=0)}{\frac{dN}{dM_i^0} 
    (Y_{j+1}^{\rm bin}-Y_{j}^{\rm bin})}
  \end{equation}
The initial distribution $dN/dM^0$ is logarithmically spaced to ensure
good statistics at the high mass end, where clusters are rare,
and to avoid creating an excessively large number of merger trees.

For the non-boosted case, the YF or yF can be found directly from the 
equilibrium relations (equations~(\ref{eq:Yintdef}) \& (\ref{eq:ymaxdef})) and
  \begin{equation} \label{eq:Yintps} 
    n_{PS}(Y,z)dY=n_{PS}(M,z)\frac{dM}{dY}dY
  \end{equation}
  \begin{equation} \label{eq:ymaxps}
    n_{PS}(y_{\rm max},z)dy_{\rm max}=n_{PS}(M,z)\frac{dM}{dy_{\rm max}}dy_{\rm max}
  \end{equation}
with $n_{PS}(M,z)$ from equation~(\ref{eq:nps}).
The derivatives are found from fits to the equilibrium
relations (see \S~\ref{sec:ynb}).

The agreement between the nonboosted merger tree-derived YFs (yFs)
and the analytic Press-Schechter YFs (yFs) is shown in Figure~\ref{fig:bnbyF}
for the flat world model.
Note that the merger trees seem to slightly overestimate the number of lower 
$Y$ or $y_{\rm max}$ (i.e. lower mass)
clusters at higher redshifts, which is due to a feature of our
merger tree procedure discussed in \S~\ref{sec:mergertrees} of this
work and \S~3.1 of \citetalias{RSR02}.

The nonboosted and boosted YFs and yFs are also compared in 
Figure~\ref{fig:bnbyF}.
The boosted YFs are almost identical to the nonboosted YFs.
The deviations from the nonboosted YFs are not systematic and are typically
of a few percent and only visible in the residual plot.
The boosted yFs, however, lie systematically above the nonboosted yFs at
all 3 redshifts considered.
The fractional deviation increases with both cluster mass and redshift.
The increase with cluster mass shows that rare events involving major
mergers of moderate mass clusters compete in frequency with the number
of rare, very massive clusters with large equilibrium values of
$y_{\rm max}$.
The increase in the bias with redshift is apparently due to the
higher merger rate in the past.
Clearly, clusters with all values of $y_{\rm max}$ are getting boosted to
higher $y_{\rm max}$ bins in the yFs over our considered range of
$y_{\rm max}$, which includes only the most massive clusters.
Such a significant and systematic bias in the yF will affect estimates
of cosmological parameters,
as discussed below in \S~\ref{sec:fityfs}.

\section{Determining Cosmological Parameters from the Merger Tree YFs and yFs}
\label{sec:fityfs}

Although mergers strongly affect the SZ signals of a small fraction of
clusters, because of the exponential high-mass drop-off in the YF and yF
the effect of mergers on cosmological model fits to these distributions
may be significant.
To quantify this effect, we derive fits based on the analytic predictions
of equations~(\ref{eq:Yintps}) and (\ref{eq:ymaxps}) to the YF and yF
using both boosted and nonboosted merger trees.
The differences between the best-fit cosmological parameters derived
in the two cases provide an estimate of the systematic bias introduced
when merging effects are neglected.

\subsection{Varying only $\Omega_M$ and $\sigma_8$} \label{sec:Omsig}

For our 3 cosmological world models, we treat the binned YFs and yFs from the
merger trees as observational data and find best-fit values for
the parameters $\Omega_M$ and $\sigma_8$ in equation~(\ref{eq:Yintps})
or (\ref{eq:ymaxps}).
Due to a near degeneracy between $\Omega_M$ and $\sigma_8$
\citep{BF98} at a single redshift, we simultaneously fit YFs and
yFs for two redshifts: at $z=0$ and at either $z=0.5$ or $z=1.0$.
While in practice SZ surveys will observe clusters in a continuous
range of redshifts, choosing only 2 redshifts simplifies the fitting
procedure and illustrates the effect of merger boosts on
these parameters.  
We choose to only fit clusters above a minimum
value of $Y^{\rm min}=10^{-5} h^{-2} {\rm Mpc}^2$ or 
$y_{\rm max}^{\rm min}=10^{-5}$.
These limits are consistent with the expected detection thresholds for
upcoming SZ surveys, such as the AMI, ACT, and SPT projects
\citep[e.g.,][]{Bar06}, and the likely confusion limit for clusters with
$M\lesssim 10^{14} h^{-1} M_\odot$ \citep{HMB07}.
These limits also keep our fits from being biased by the large number
of clusters at low masses.

To evaluate the extent to which merger boosts affect the estimation of
$\Omega_M$ and $\sigma_8$, we compare their fitted values from the
boosted YFs and yFs to the fitted values from the nonboosted  
YFs and yFs.
We do not compare best-fit parameters 
to the values used to create the merger trees because
the trees tend to slightly overestimate the MF, an effect which increases
with redshift and is discussed in \S~\ref{sec:mergertrees}.
However, since we are only interested in {\it relative} changes to the
YF or yF due to boosts in $Y$ or $y_{\rm max}$, this bias in the MF
does not affect our results, though the best-fit parameters found from 
the nonboosted YFs or yFs may differ from the parameter
values used to create the trees.
Also, any bias caused by our chosen fitting method is accounted 
for by directly comparing the two YFs or yFs.

The best-fit values of $\Omega_M$ and $\sigma_8$ for the flat, open,
and EdS cosmological world models are given in Table~\ref{tab:omsig8}
for both $Y$ and $y_{\rm max}$.
The parameter values used to create the merger trees are summarized
in Table~\ref{tab:omsig8_true} for reference.
In general, the results are independent of world model; cosmological 
parameter fits tend to be biased in the same direction by about the 
same amount.
However, boosts to $Y$ have almost no effect 
on fits to $\Omega_M$ and $\sigma_8$;
the changes due to mergers are generally less than 1\%
and are not clearly systematic.

In contrast, boosts to $y_{\rm max}$ significantly bias the values of 
these parameters:
$\Omega_M$ is underestimated by 15-30\% and
$\sigma_8$ is overestimated by 10-20\%.
The main effect of merger boosts is to increase the number of clusters 
detected in a particular ${\rm yF}_j$ bin;
in other words, there is a systematic increase in the yF, as shown in the right
panel of Figure~\ref{fig:bnbyF}.
An overall increase in the normalization of the yF leads to an
increase in the normalization of the spectrum of initial density
perturbations, $\sigma_8$.
The total matter content, $\Omega_M$, is also sensitive to the 
normalization, but it is nearly degenerate with
$\sigma_8 \approx 0.6\Omega_M^{-1/2}$
\citep{BF98} at a given single redshift.
However, $\Omega_M$ is more sensitive to the change in the yF over time --
the greater the density of matter, the faster structure will grow.
If various cosmologies with nearly identical yFs at $z=0$
are considered, those cosmologies with smaller values of $\Omega_M$
(and thus larger values of $\sigma_8$) would produce yFs at $z>0$
that lie above the yFs of cosmologies with larger $\Omega_M$ values.
As described in \S~\ref{sec:yfs}, merger boosts raise the yF 
most strongly
at higher redshifts, so the change in the yF from one redshift
to another is smaller than for nonboosted yFs, indicating a slower
structure growth rate and therefore a smaller $\Omega_M$.
The overall effect of mergers seems to vary with redshift; $\Omega_M$
and $\sigma_8$ are found to be less biased when utilizing the yF at 
higher redshift ($z=1$) even though this yF is fractionally more biased
than the yFs at $z=0$ or $z=0.5$.

\subsection{Fitting the Dark Energy Equation of State 
Parameter $w$} \label{sec:w}

Clusters of galaxies have been used to constrain the equation of
state parameter $w$ of dark energy,
and there are extensive plans to improve these measurements in the future
using SZ surveys
\citep[e.g.,][]{HMH01, WBK02}.  
In the $\Lambda$CDM flat world model, dark energy is assumed to 
take the form of a cosmological constant, which has a fixed $w=-1$.
Here, we assess the effect of mergers on the determination of
$w$ by allowing $w$ to vary along with $\Omega_M$ and
$\sigma_8$ in fits to the flat world model YFs and yFs,
following the same procedure outlined in \S~\ref{sec:Omsig}.  
We need new
analytic, nonboosted Press-Schechter YFs and yFs that incorporate 
$w\ne-1$, which we write as
$n_{PS}(Y,z,w)$ and $n_{PS}(y_{\rm max},z,w)$.  
The same basic
form of $n_{PS}$ can be generalized to a constant $w \ne -1$ 
and a slowly varying parameterization of 
$w(z) = w_0+w_1a(1-a) = w_0+w_1z/(1+z)^2$, where $a$ is the scale
factor and $w_0$ and $w_1$ are constants and $w_1$ is small.
In a flat ($\Omega_M+\Omega_{DE}=1$) universe, we change the
expression for the growth function $D(z)$ as given in Appendix A
of \citetalias{RSR02} and correct the comoving volume element $dV$
such that
  \begin{equation} \label{eq:wvol}
    \left( \frac{dN}{dVdY} \right)^{w} =
    \left( \frac{dN}{dVdY} \right)^{w=-1, \delta_c(D(z,w))}
    \left( \frac{d_{\rm A}^{w=-1}}{d_{\rm A}^{w}} \right)^2
    \left( \frac{dV^{w=-1}}{dV^{w}} \right)
    \, ,
  \end{equation}
where $\left(\frac{dN}{dVdY}\right)^{w=-1,\delta_c(D(z,w))}$
is $n_{PS}(Y,z)$ for the $\Lambda$CDM cosmology, but with
the critical overdensity $\delta_c$ given by the new growth function
$D(z,w)$
[equation~(20) from \citet{Per05} for constant $w$
or
equation~(14) from \citet{WS98} for for $w(z)$], and 
$d_{\rm A}$ is the angular diameter distance.
The ratio of volumes is
  \begin{equation} \label{eq:vol}
    \left( \frac{dV^{w=-1}}{dV^{w}} \right) =
    \left( \frac{d_{\rm A}^{w=-1}}{d_{\rm A}^{w}} \right)^2
    \,
    \left[ \frac{E(z,w)}{E(z,w=-1)} \right]
    \, ,
  \end{equation}
where $E(z,w)=[\Omega_M(1+z)^3+\Omega_{DE}(1+z)^{3+3w}]^{1/2}$.
The same expression applies for the yFs by replacing
$Y$ with $y_{\rm max}$ and dropping the factor
$\left( \frac{d_{\rm A}^{w=-1}}{d_{\rm A}^{w}} \right)^2$
from equation~(\ref{eq:wvol}).

\subsubsection{Constant $w\ne -1$}\label{sec:wconst}

When we allow for constant values of $w$ that are not necessarily equal 
to $-1$, we find results qualitatively similar to what was found 
previously when only $\Omega_M$ and $\sigma_8$ were varied.
Again, the boosted YFs give back nearly identical
values for all 3 parameters to within $\lesssim 1\%$.
For $y_{\rm max}$,
merger boosts are found to
bias the fitted values for $\Omega_M$ and $\sigma_8$ even more
strongly, underestimating $\Omega_M$ by 30-45\% and overestimating
$\sigma_8$ by 20-25\%.
Also, $w$ is found to be more negative in the boosted yFs by
25-45\%, making $y_{\rm max}$ a poor proxy if one aims to constrain
the nature of dark energy.
These results are summarized in Table~\ref{tab:flatw}.

In the case of $y_{\rm max}$, the boosted yFs favor more negative values
of $w$ due to $w$'s impact on structure formation.
The yF is overestimated to a greater extent at larger
redshifts (see Figure~\ref{fig:bnbyF}),
which mimics more structure in the recent ($z\lesssim 1$) past.
In turn, the appearance of more collapsed structures in the past
relative to the present time implies that recent structure formation was
slower than it actually has been,
and that structure formation in the far past was
correspondingly faster.
In general, if we compare the effect of different values of $w$ 
on structure formation by holding the present yF fixed, a more negative
$w$ is better able to slow down cluster formation at later times
as the strength of dark energy grows with the scale factor $a$ since
$\Omega_{DE} = \Omega_{DE,0} (1+z)^{3(1+w)}/E^2(z)$.
If cluster formation is slowing at the current epoch, when dark energy has
recently become dominant, there must be more clusters in the recent past
compared to the yF of clusters under the influence of a less negative $w$.

A more negative $w$ allows for even smaller values of $\Omega_M$ to be
fit to the boosted yFs, compared to its best-fit values
when only $\Omega_M$ and $\sigma_8$ are varied.
By anchoring the current yF, a more negative $w$ {\it decreases} the
influence of dark energy in the past, so $\Omega_M$ does not need to
be as large to form the same amount of structure.
The dark energy equation of state does not as directly affect the 
overall normalization of the yFs, so the bias to $\sigma_8$
remains consistent with the fixed $w=-1$ fits.

\subsubsection{Slowly Varying $w(z)$}\label{sec:wa}

If dark energy is not due to a cosmological constant, then it is
possible that its equation of state might vary.
We have also determined the effect that merger boosts can have on
the SZ determination of the evolution of dark energy.
We only consider the effect of boosted YFs in this section, 
due to the difficulty of using yFs to pin down even constant values
of $w$.
Choosing the parameterization of $w = w_0 + w_1z/(1+z)^2$, where $w_0$
and $w_1$ are constants, we determined $\Omega_M$, $\sigma_8$, $w_0$, and
$w_1$ by fitting the boosted and nonboosted YFs.  
The validity of the form of the growth function we use 
for a flat universe requires that
$\vert \frac{dw}{d\Omega_M} \vert \ll \frac{1}{1-\Omega_M}$, which implies
that $w_1 \ll 1$ for $w_0 \approx -1$ and
$\Omega_M \approx 0.3$ \citep{WS98}.
We do not constrain our best-fit value of $w_1$ according to this
requirement, however, nor do we consider any other parameterization
of $w$.

We found that $w_1$ was not well-constrained by fitting the YFs or yFs
at only two redshifts.
Thus, we simultaneously fit the distribution functions at the three
redshifts $z=0$, 0.5, and 1.
As in the constant $w$ case, the boost-derived values of $\Omega_M$,
$\sigma_8$, and $w_0$ deviated from the nonboosted values 
only slightly, by $+0.5\%$, $-0.2\%$, and +2\%, respectively.
The best-fit values of $w_1$ increased by 0.15 from the nonboosted value
of $-0.19$ to a value for the boosted YF of $-0.04$.
For the assumed variation of $w$ with $z$, the largest change in $w$
occurs between the present time ($z=0$) and  $z=1$;
that change is $\Delta w = w_1 / 4$.
Thus, the merger boost effects on YF alter the maximum change in the $w$
by about 0.04.
This is about twice as large as the effect on $w_0$, but is still
relatively small.

\section{Discussion and Summary} \label{sec:concl}

We have determined the effects of cluster mergers on their SZ properties,
particularly the integrated $Y$ and maximum $y_{\rm max}$
Comptonization parameters.
From a set of hydrodynamical/N-body simulations of cluster mergers,
we determined the evolution of $Y$ and $y_{\rm max}$ 
over the period of interaction for mergers of various mass ratios and 
impact parameters, and we found that mergers temporarily ``boost" both
$Y$ and $y_{\rm max}$.
For $y_{\rm max}$, the boosts can be as large as an order of magnitude,
although they occur for a short time (typically about half the sound
crossing time of the cluster), with the largest boosts occurring
near the time of first core crossing.
For major mergers,
the boosts in the maximum Comptonization parameter generally exceed the
increase in $y_{\rm max}$ when the systems have come into equilibrium.

On the other hand, the boosts in $Y$ are smaller (less than a factor
of two), although they last longer (about two sound crossing times).
Most importantly, the boosts in $Y$ for major mergers are smaller than the
increases in $Y$ when the merged clusters have come into equilibrium.
Thus, one can think of the merger ``boost'' in $Y$ as representing a
stage in the evolution from two separate equilibrium values to the
final merged value, and not really being a ``boost'' at all.
A simple physical argument explains why the transient
boosted values of $Y$ are smaller than the final equilibrium values.
From equation~(\ref{eq:Y}), it follows that $Y$ is just proportional
to the total thermal energy content of the electrons in  the clusters,
or just the total thermal energy if the electrons and ions are in
equipartition.
Now, a cluster merger involves the conversion of the bulk kinetic energy
of the merging clusters into thermal energy.
When the merger is complete, there is very little bulk kinetic energy
remaining (perhaps, weak rotation or turbulence).
Thus, one expects the thermal energy content of the merging clusters to
be largest when they have achieved (or nearly achieved) equilibrium.
Thus, the final equilibrium value of $Y$ will tend to be larger than any
transient value during the merger.

We generalized the SZ boosts
to mergers of arbitrary mass ratio and impact parameter
and traced the merger, and thus boost, history of clusters with redshift
using the EPS merger tree formalism.
In general, merger boosts induced a relatively small scatter, $\sim2\%$,
{\it below} the equilibrium $Y$--$M$ relation, 
while mergers induced a large scatter of 25-30\% above the
$y_{\rm max}$--$M$ equilibrium relation.
Power-law fits to $Y$ and $y_{\rm max}$ as a function of mass
show that while boosts do not affect the slope of the fit,
the normalization was lowered by $<1\%$ for $Y$ and
raised by $\sim 10\%$ for $y_{\rm max}$.

We used the merger trees to derive the distribution functions of SZ
parameters, YF and yF.
We found that the boosted YF was not significantly biased relative to 
the nonboosted YF, 
while the boosted yF was strongly biased
above the nonboosted yF for all redshifts.
In general, the size of the merger-induced bias increased with 
redshift and with cluster mass.

Using the YFs and yFs, we determined the best-fit values 
for the cosmological parameters
$\Omega_M$ and $\sigma_8$ for the flat, open, and EdS world models,
and also the dark energy equation of state parameter $w$ for the flat
universe.
Comparing the best-fit values of $\Omega_M$ and $\sigma_8$ for
the nonboosted and boosted YF, no significant difference ($<1\%$) was
observed.
In contrast, the boosts to the yF decreased the best-fit value of 
$\Omega_M$ by 15-30\% for the flat and open world models and increased 
the best-fit value of $\sigma_8$ by 10-20\% for all world models.
These results stem mainly from an overall increase in the yFs, which
pushes $\sigma_8$ to larger values, and a greater increase in the
boosted yF at higher redshifts relative to lower redshifts, which pushes 
$\Omega_M$ to smaller values.
Allowing for a constant $w \ne -1$ in the flat world model,
no systematic difference in fitted cosmological
parameters was found between the two sets of YFs, though the merger-induced bias
to $\Omega_M$, $\sigma_8$, and $w$ was exacerbated when using the yFs.
We also considered a time-varying $w(z)$ for the YFs, for which
$\Omega_M$, $\sigma_8$, and $w_0$ were found to be consistent with the 
previous results for a constant $w \ne -1$, though
boosts increased the best-fit value of the dark energy evolution
parameter $w_1$ by about 0.15.
The largest change in $w$ occurs between $z=1$ and $z=0$ in this model;
thus, the change in $w$ might be affected by as much as 0.04.
This is about twice as large as the maximum change in the present-day
value of $w_0$, but still is relatively moderate.

These results agree with previous work which indicates that 
global observables such as $Y$ or
the equivalent X-ray/mass proxy $Y_X$ are more robust as mass proxies than 
the central or maximum Comptonization parameter.
For example, from semi-analytic models of the intracluster medium 
(ICM), \citet{RS06} 
generally find that $Y \propto f_g M^{5/3}$, equivalent to our
equilibrium definition of $Y$, with only a small scatter due
to internal physics.
A number of studies have used cosmological N-body simulations and re-simulated
forming clusters with various kinds of gas physics
to evaluate the scatter in the $y$--$M$ 
relations \citep{Nag06, BTD+07}.
It is generally found that the normalization $A$ varies significantly 
depending on the ICM physics, though the slope $\alpha$ does not.
\citet{Nag06} reports a scatter of 10-15\% in the $Y$--$M$ relation,
much larger than our scatter of $\sim 2$-3\%.
Also, \citet{KVN06} defines an X-ray observable $Y_X = T_X M_{\rm gas,500}$,
which is similar to our $Y \propto \int n_e T_e dV \sim T_e \int n_e dV
\propto T_e M_{\rm gas}$; they find a scatter in the relation of 5\%-7\%.

While these studies intrinsically include mergers, they have limited
statistics as they generally consist only of a small number of systems,
$\sim10$ or so.
Some studies have considered somewhat larger cluster simulation samples
including
hundreds of clusters \citep{MHB+05, DKL+04} from cosmological 
simulations.
Our results agree with their conclusions that the 
$Y$--$M$ relation is relatively stable to mergers, unlike the
$y_{\rm max}$--$M$ relation. 
\citet{MHB+05} find a scatter in their $Y$--$M$ relation of 3-4\%
and in their $y_{\rm max}$--$M$ relation
of $\sim17\%$ at $z=0$ due to mergers and other ICM physical
processes.
These results compare well with our scatter of 2\% and 24\%,
respectively.

\begin{figure}
\vskip2.3truein
\includegraphics{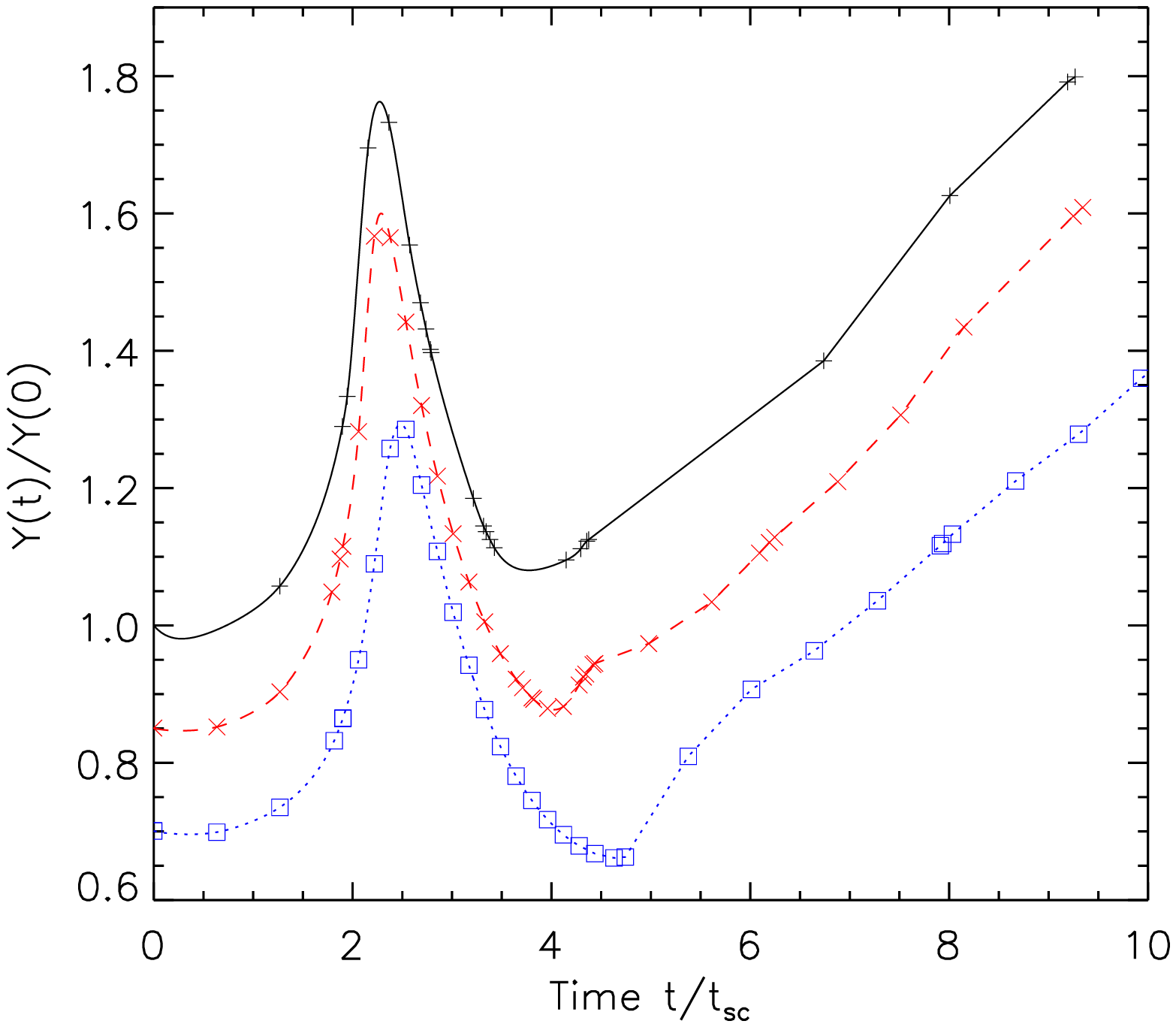}
\includegraphics{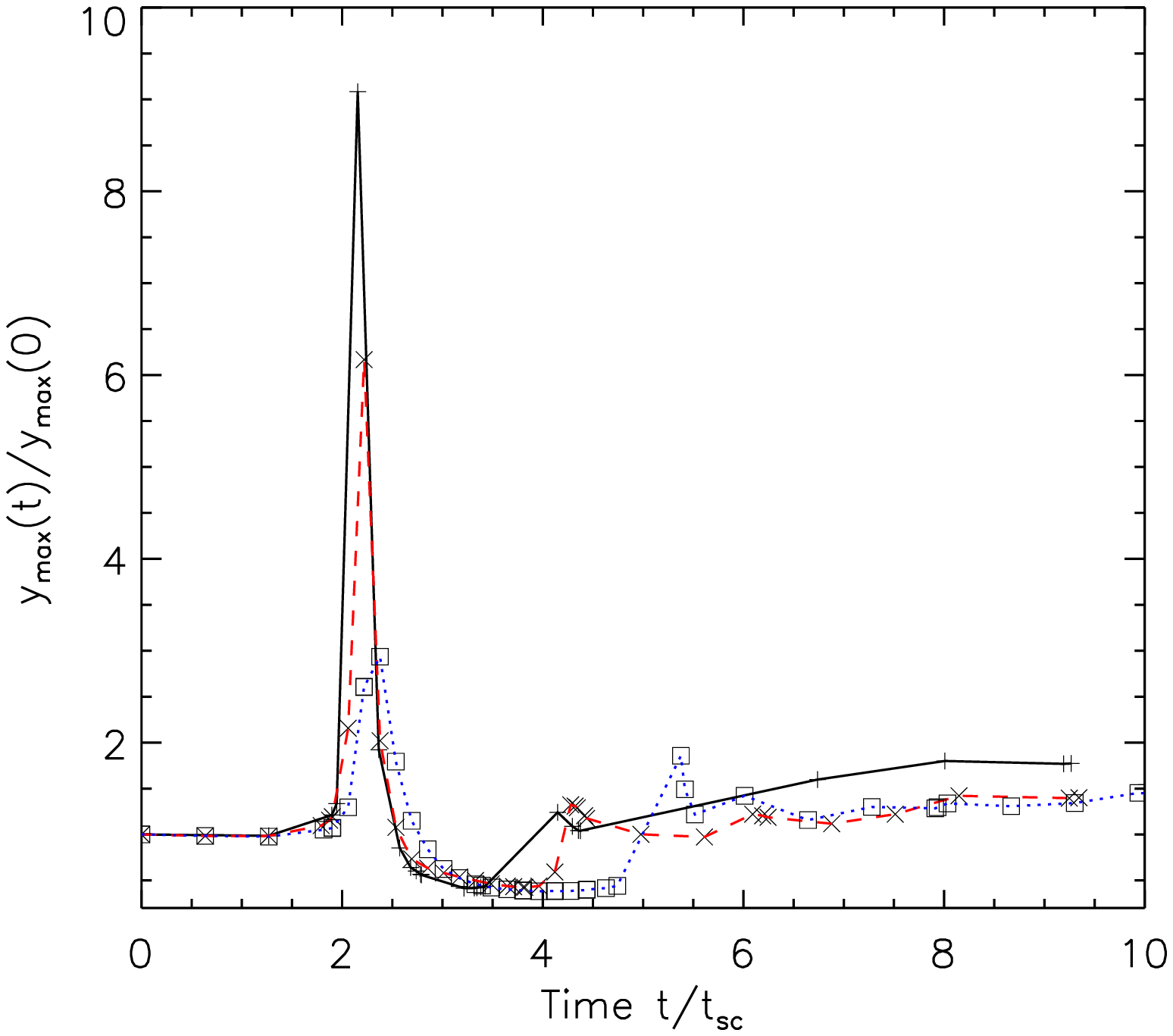}
\vskip2.3truein
\includegraphics{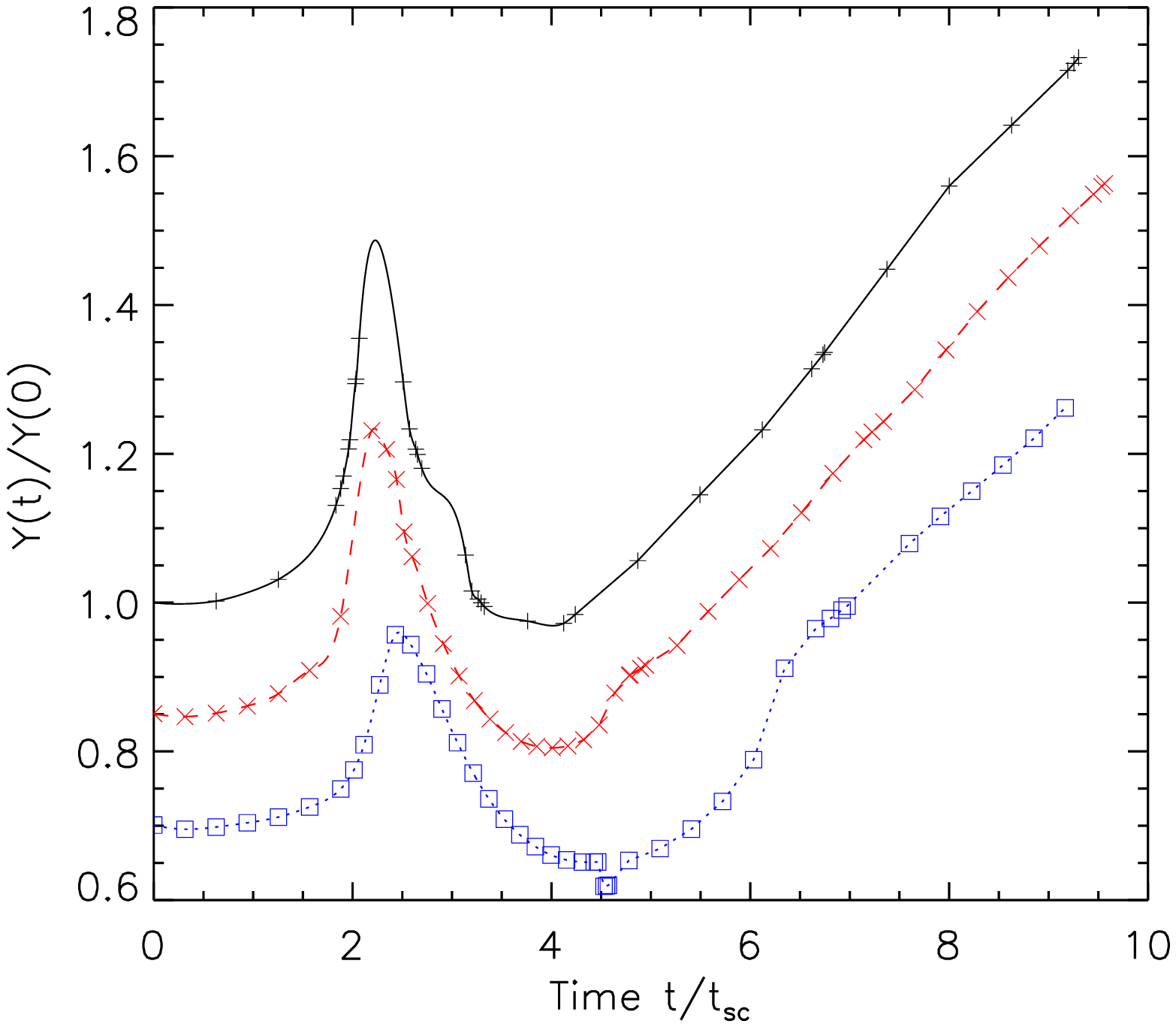}
\includegraphics{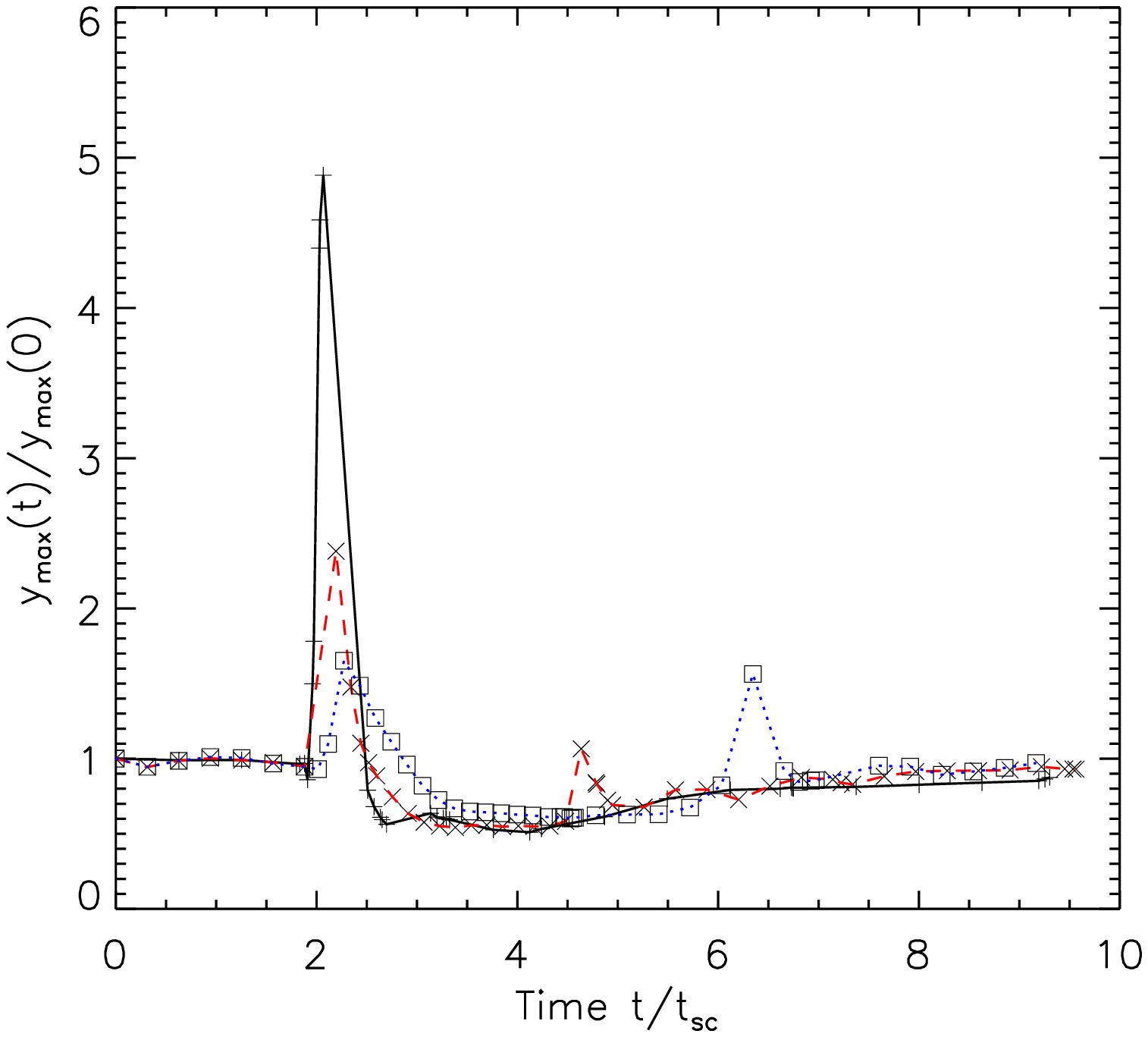}
\vskip2.3truein
\includegraphics{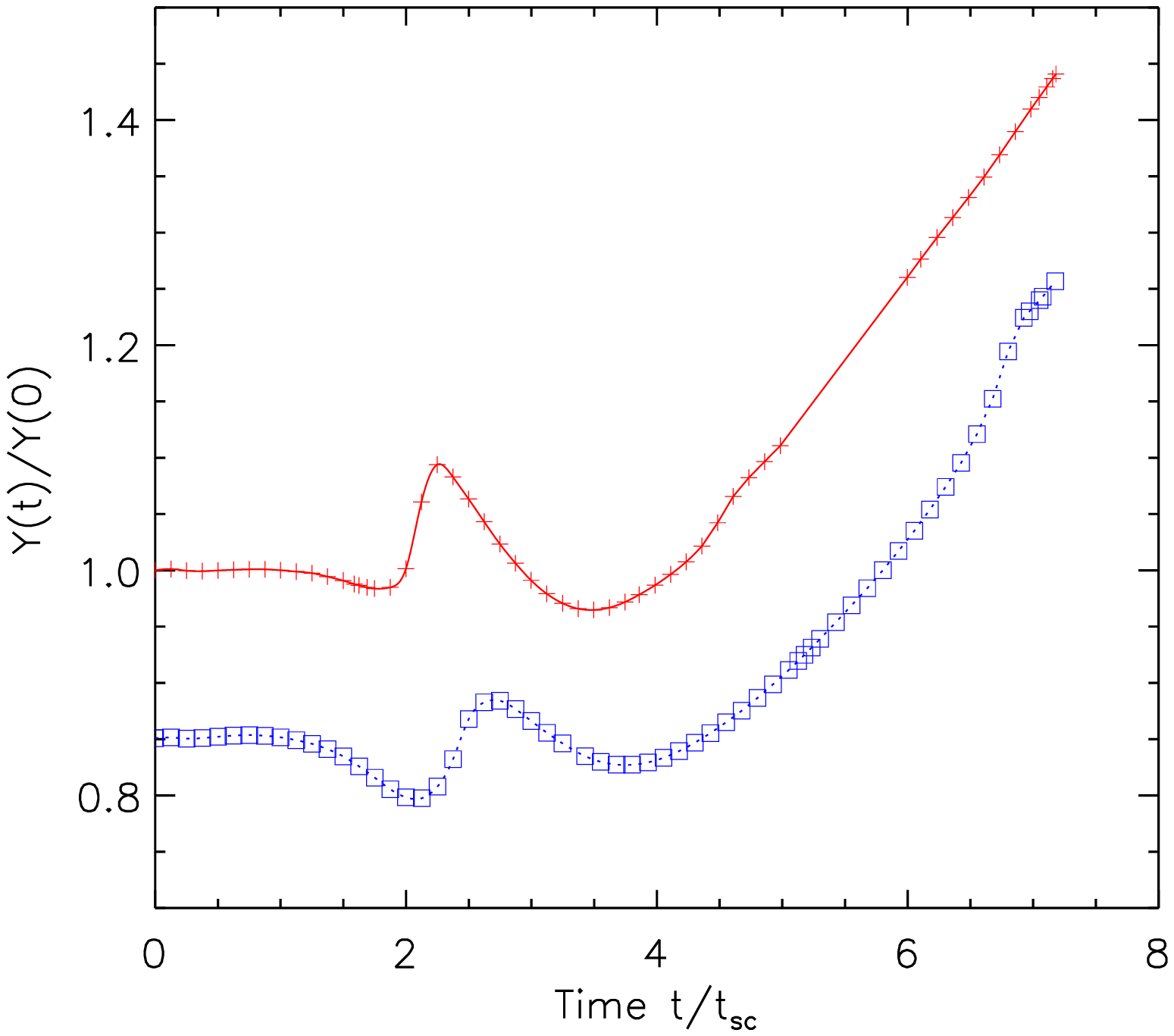}
\includegraphics{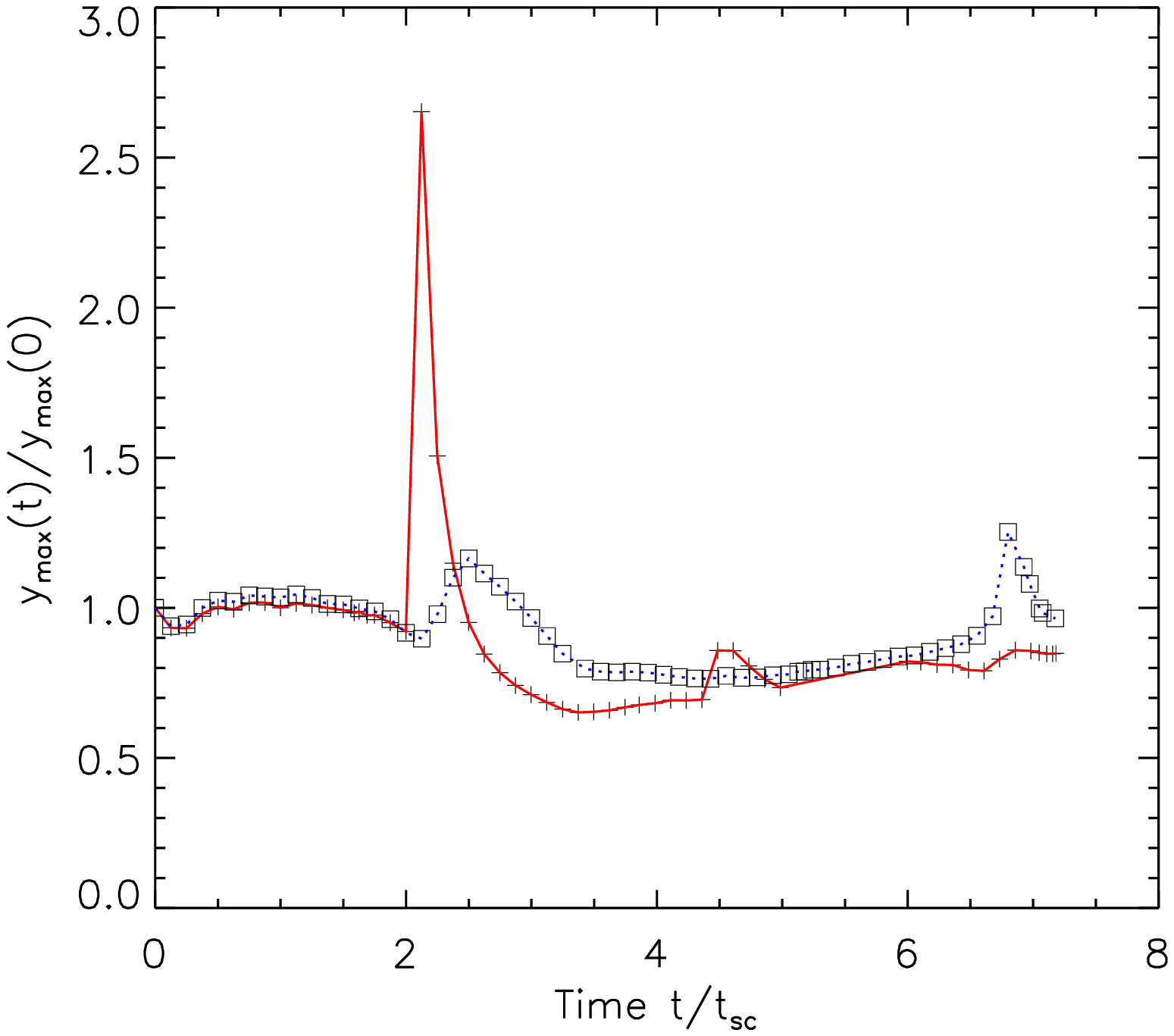}
\caption{Evolution of the SZ effect during a merger.  
In each panel, different curves are for different values of the merger
impact parameter: $b=0$ (solid, black line), 2 $r_s$ (dashed,red line), 
and 5 $r_s$ (dotted, blue line),
where $r_s$ is the NFW scale radius of the more massive cluster.
In the left panels, the $b = 2 r_s$ simulation run is offset downward by 0.15
and the 5$r_s$ run is offset downward by 0.3 for clarity.
The time is scaled by the sound crossing time $t_{\rm sc}$
of the more massive premerger cluster.
{\it Left:} Integrated Comptonization parameter $Y$ versus time
for the 1:1 (top), 1:3 (middle), and 1:6.5 (bottom) mass ratios.
{\it Right:} Maximum Comptonization parameter $y_{\rm max}$ versus time
for the 1:1 (top), 1:3 (middle), and 1:6.5 (bottom) mass ratios.
The mergers are observed 90$^\circ$ to the merger axis and
in the merger plane.
\label{fig:yt}}
\end{figure}

In a study similar to this work, \citet{PBM+07} take a suite of binary cluster
merger simulations to assess the effect of various observables,
including SZ parameters, on scaling relations during mergers.
The evolution of $y_{\rm max}$ (which they call $y_o$) in their
simulations is qualitatively similar to our results in Figure~\ref{fig:yt}
for various impact parameters and mass ratios.
They also consider an integrated Comptonization parameter, but it is
only integrated out to a radius $r_{2500}$ and is thus much more dominated by
core effects and not equivalent to our $Y$, which is effectively integrated
to at least $r_{200}$, the virial radius.

The large number of galaxy clusters expected from upcoming SZ surveys,
both locally and at potentially high redshifts, heightens the prospects
that clusters could play a decisive role in the era of precision cosmology,
especially if the robustness of $Y$ as a proxy for mass
is confirmed in real cluster samples.

\acknowledgments
This work was supported in part by the National Aeronautics and
Space Administration through {\it Chandra} awards
GO5-6126X,
and
TM7-8010X,
and through {\it XMM-Newton} awards
NNG05GO50G,
NNG06GD54G,
NNX06AE76G,
and
NNX06AE75G,
and through {\it Suzaku} awards
NNX06AI37G,
and
NNX06AI44G.
DRW acknowledges the support of a Virginia Space Grant Consortium
graduate fellowship.
We would like to thank the referee for very helpful comments.

% \clearpage

% Tables

\begin{deluxetable}{lccc}
\tablecaption{Cosmological Parameter Values Used to Create Merger Trees
\label{tab:omsig8_true}}
\tablewidth{0pt}
\tablehead{ Model & $\Omega_M$ & $\Omega_{\Lambda}$ & $\sigma_8$}
\startdata
Flat & 0.3 & 0.7 & 0.834 \\
Open & 0.3 & 0.0 & 0.827 \\
EdS  & 1.0 & 0.0 & 0.514 \\
\enddata
\end{deluxetable}

\begin{deluxetable}{lcccccccc}
\tablecaption{Merger-Induced SZ--$M$ Relations and Scatter
\label{tab:scatter}}
\tablewidth{0pt}
\tablehead{ & Model & $z$ & $\sigma_{\rm eq}$ & $\sigma_{\rm fit}$ & 
$A_{\rm nb} $ & $A_{\rm b} $ & $\alpha_{\rm nb}$ & 
$\alpha_{\rm b}$}
\startdata
$Y$ & Flat &  0.0 & 0.0205 & 0.0218 & 12.0 & 11.9 & 1.91 & 1.91 \\
    &      &  0.5 & 0.0190 & 0.0205 &      & 11.9 &      & 1.91 \\
    &      &  1.0 & 0.0197 & 0.0212 &      & 11.9 &      & 1.91 \\
    & Open &  0.0 & 0.0207 & 0.0220 & 12.0 & 11.9 & 1.91 & 1.91 \\
    &      &  0.5 & 0.0191 & 0.0207 &      & 11.9 &      & 1.91 \\
    &      &  1.0 & 0.0203 & 0.0216 &      & 11.9 &      & 1.91 \\
    & EdS  &  0.0 & 0.0214 & 0.0228 & 12.0 & 11.9 & 1.91 & 1.91 \\
    &      &  0.5 & 0.0194 & 0.0209 &      & 11.9 &      & 1.91 \\
    &      &  1.0 & 0.0222 & 0.0235 &      & 11.9 &      & 1.91 \\
$y_{\rm max}$ & Flat & 0.0 & 0.292 & 0.241 & 8.25 & 9.01 & 1.24 & 1.26 \\
              &      & 0.5 & 0.361 & 0.271 &      & 9.39 &      & 1.27 \\
              &      & 1.0 & 0.412 & 0.289 &      & 9.70 &      & 1.27 \\
              & Open & 0.0 & 0.293 & 0.246 & 8.25 & 8.96 & 1.24 & 1.26 \\
              &      & 0.5 & 0.329 & 0.256 &      & 9.20 &      & 1.27 \\
              &      & 1.0 & 0.375 & 0.279 &      & 9.46 &      & 1.26 \\
              & EdS  & 0.0 & 0.414 & 0.290 & 8.25 & 9.72 & 1.24 & 1.27 \\
              &      & 0.5 & 0.485 & 0.301 &      & 10.3 &      & 1.27 \\
              &      & 1.0 & 0.531 & 0.299 &      & 10.7 &      & 1.28 \\
\enddata
\end{deluxetable}

\begin{deluxetable}{lccc|cccc}
\tablecaption{Best-Fit Values for $\Omega_M$ and $\sigma_8$ for
Three World Models
\label{tab:omsig8}}
\tablewidth{0pt}
\tablehead{ & Model & $z$ & Boosts? & $\Omega_M$ &
Difference & $\sigma_8$ & Difference}
\startdata
$Y$ & Flat & 0,0.5 & no & 0.287  &         &  0.857 & \\
 & &      & yes &         0.289  &  0.7\%  &  0.854 &  -0.4\% \\
 &      &  0,1.0 & no &   0.277  &         &  0.865 & \\
 & &      & yes &         0.277  &  0.0\%  &  0.865 &  0.0\% \\
    & Open & 0,0.5 & no & 0.278  &         &  0.857 & \\
 & &      & yes &         0.279  &  0.4\%  &  0.856 &  -0.1\% \\
 &      &  0,1.0 & no &   0.279  &         &  0.855 & \\
 &      & & yes &         0.280  &  0.4\%  &  0.855 &  0.0\% \\
    & EdS & 0,0.5 & no & 0.932  &         &  0.531 & \\
 &    &  & yes &         0.931  & -0.1\%  &  0.531 &  0.0\% \\
 &    &   0,1.0 & no &   0.874  &         &  0.541 & \\
 &    &  & yes &         0.873  & -0.1\%  &  0.541 &  0.0\% \\
\hline
$y_{max}$ & Flat & 0,0.5 & no &  0.295 &         &   0.844 & \\
 &      & & yes &                0.199 &   -33\% &   1.020 &    21\% \\
 &      & 0,1.0 & no &           0.267 &         &   0.870 & \\
 &      & & yes &                0.229 &   -14\% &   0.976 &    12\% \\
          & Open & 0,0.5 & no &  0.282 &         &   0.848 & \\
 &      & & yes &                0.213 &   -24\% &   0.984 &    16\% \\
 &      & 0,1.0 & no &           0.281 &         &   0.848 & \\
 &      & & yes &                0.236 &   -16\% &   0.954 &    13\% \\
          & EdS & 0,0.5 & no &  0.953 &          &   0.524 & \\
 &    &  & yes &                0.921 &   -3.4\% &   0.589 &    12\% \\
 &    &  0,1.0 & no &           0.905 &          &   0.532 & \\
 &    &  & yes &                0.924 &    2.1\% &   0.590 &    11\% \\
\enddata
\end{deluxetable}

\begin{deluxetable}{lcc|cccccc}
\tablecaption{Best-Fit Flat World Models with Constant $w$ \label{tab:flatw}}
\tablewidth{0pt}
\tablehead{ & $z$ & Boosts? & $\Omega_M$ &
Difference & $\sigma_8$ & Difference & $w$ & Difference}
\startdata
$Y$ & 0,0.5 & no & 0.314  &         &  0.837 &         & -0.879 & \\
 & & yes &         0.316  &  0.6\%  &  0.835 &  -0.2\% & -0.885 & 0.7\% \\
 &  0,1.0 & no &   0.275  &         &  0.874 &         & -1.062 & \\
 & & yes &         0.271  &  -1.5\% &  0.877 &  0.3\%  & -1.080 & 1.7\% \\
\hline
$y_{max}$ & 0,0.5 & no &  0.324 &         &   0.823 &         & -0.861 & \\
 & & yes &                0.173 &  -47\%  &   1.082 &    24\% & -1.255 & 46\% \\
 & 0,1.0 & no &           0.279 &         &   0.863 &         & -0.987 & \\
 & & yes &                0.192 &  -31\%  &   1.045 &    21\% & -1.240 & 26\% \\
\enddata
\end{deluxetable}

% \clearpage

% Figures

\begin{figure}
\plottwo{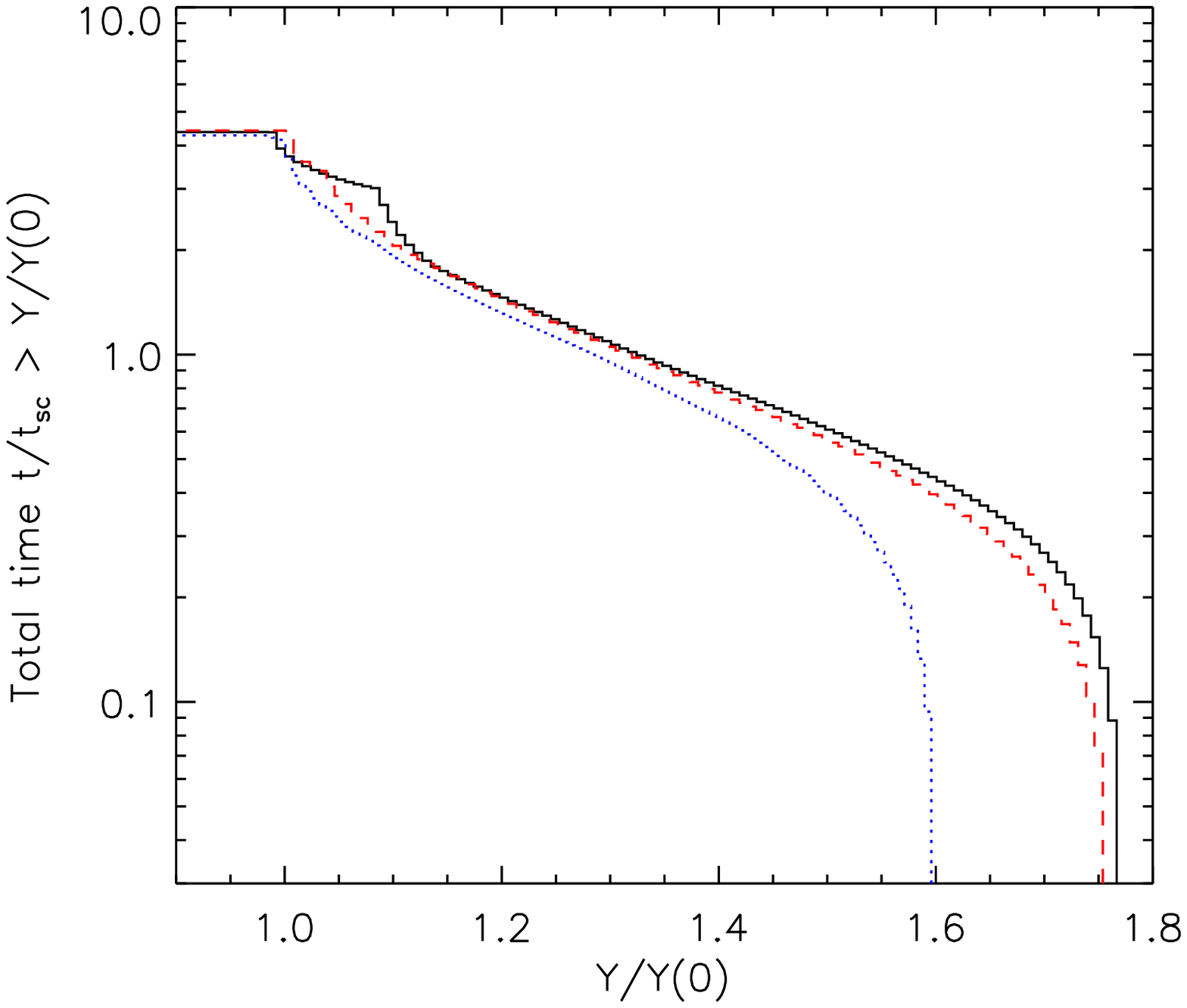}{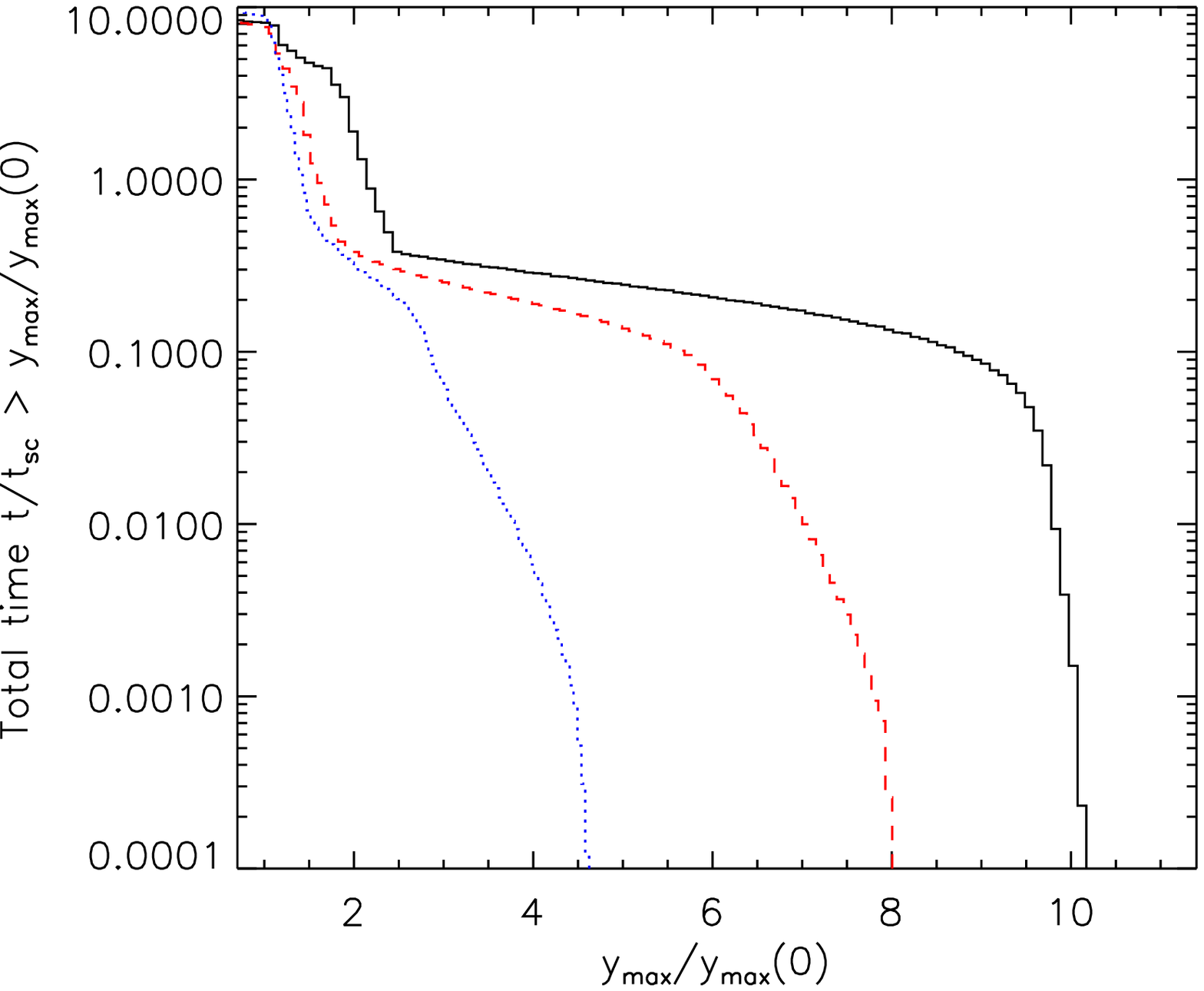}
\caption{{\it Left:} Histogram of the total time the integrated
Comptonization parameter
$Y$ is above some fraction of its initial premerger value $Y(0)$, scaled by
the sound crossing time $t_{\rm sc}$ of
the more massive premerger cluster.
Histograms are shown for equal-mass mergers at 3 impact parameters $b=0$ 
({\it solid, black line}), 2 $r_s$ ({\it dashed, red line}), 
5 $r_s$ ({\it dotted, blue line}),
where $r_s$ is the NFW scale radius of the more massive cluster.
{\it Right:} Histogram of times for $y_{\rm max}$.
\label{fig:histy}}
\end{figure}

\begin{figure}
\plotone{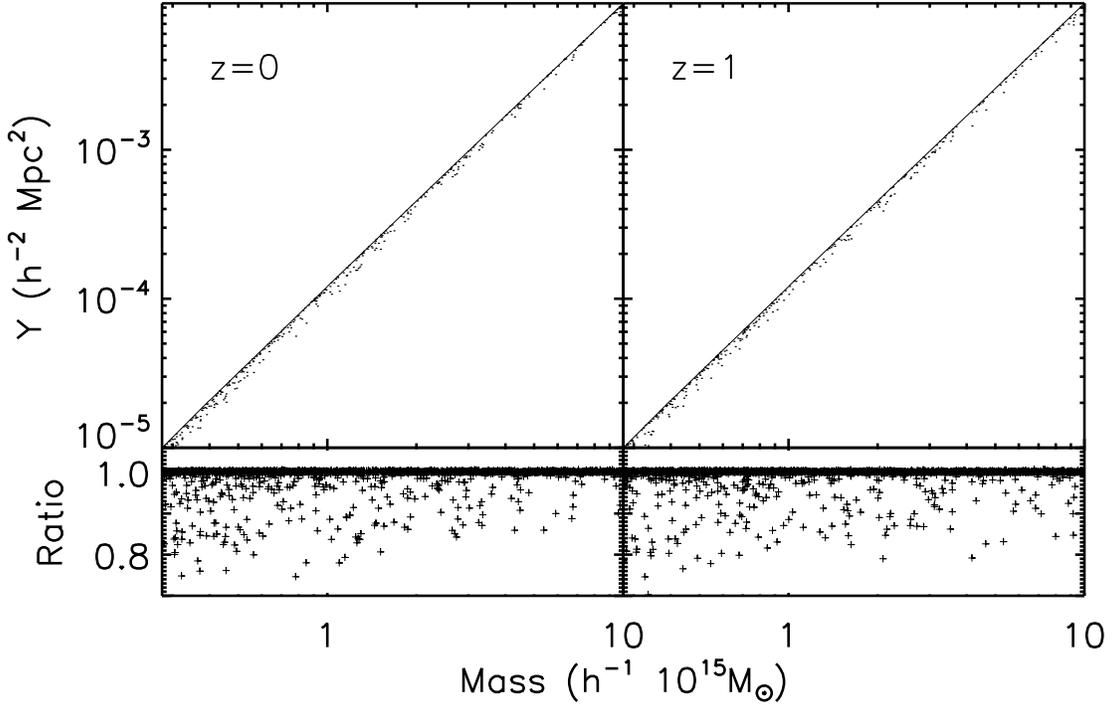}
\caption{Integrated Comptonization parameter $Y$
({\it top panels})
versus total mass in the flat cosmology at $z=0$ ({\it left panels})
and $z=1$ ({\it right panels}) for clusters with
$Y > 10^{-5} h^{-2} \, {\rm Mpc}^2$.
The combined mass of both merging clusters is used if $Y$
is boosted or
$t_{\rm obs}>t_{\rm merge}$, where $t_{\rm merge}$ is the time of maximum boost.
The apparent solid line is the result of many individual clusters
at or near their equilibrium values of $Y$.
In the {\it bottom panels}, the ratio of the boosted clusters to their
equilibrium values for each redshift is shown.
Each panel contains 5190 clusters.
\label{fig:yiscat}}
\end{figure}

\begin{figure}
\plotone{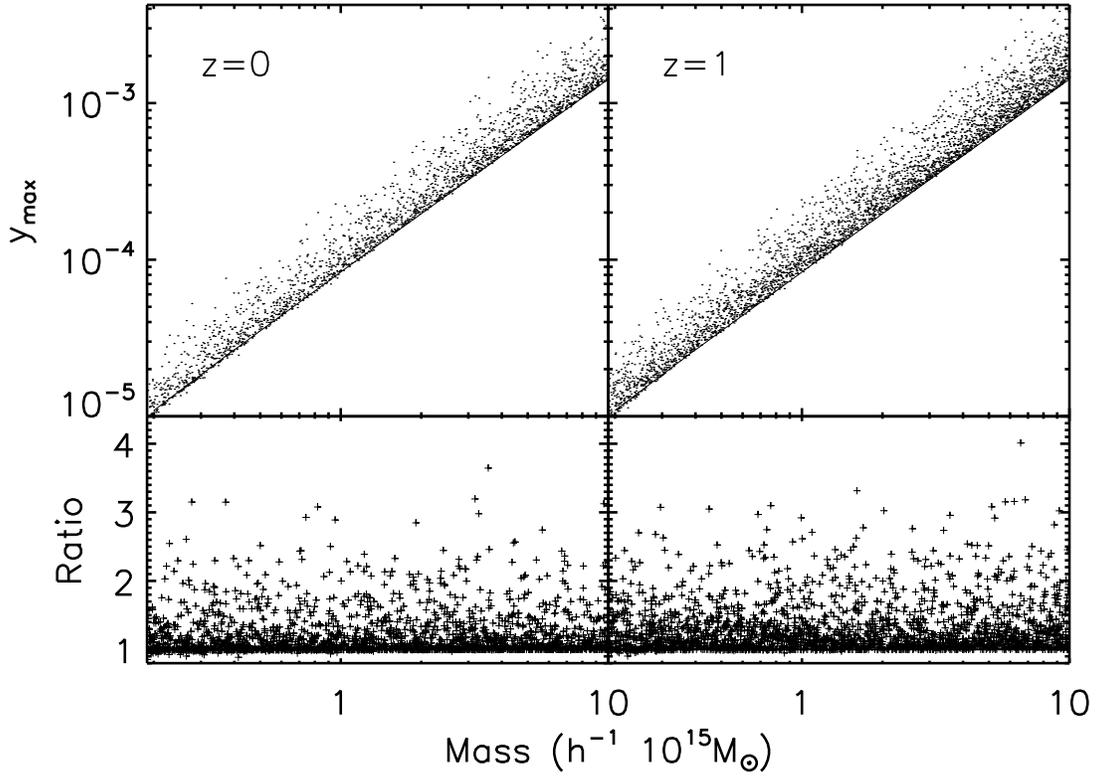}
\caption{Same as Figure~\ref{fig:yiscat}, but for the maximum Comptonization
parameter $y_{\rm max}$, for clusters with $y_{\rm max} > 10^{-5}$.
Each panel contains 5663 clusters.
\label{fig:ymscat}}
\end{figure}

\begin{figure}
\plottwo{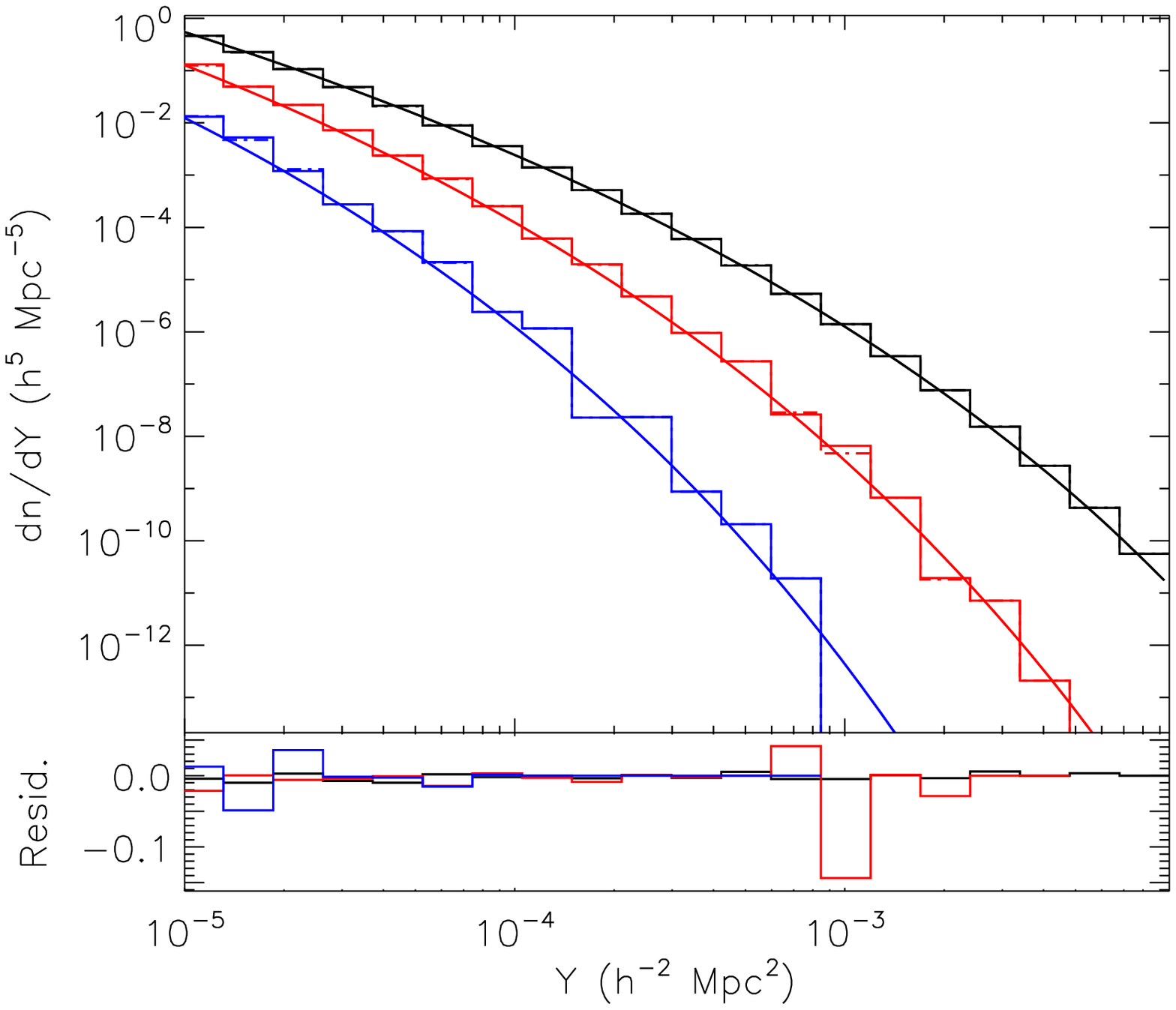}{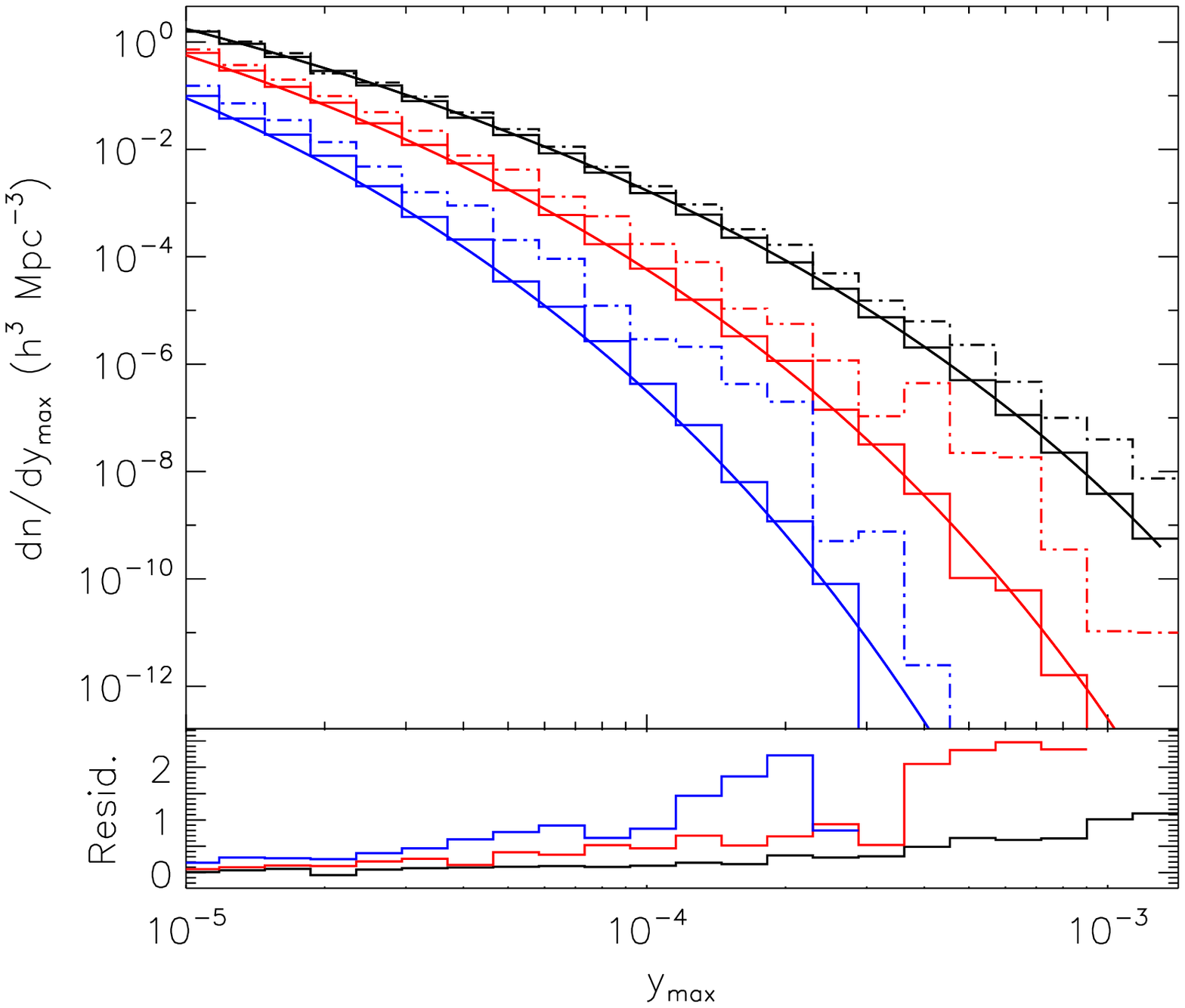}
\caption{Boosted ({\it dashed line}) versus nonboosted ({\it solid line}) 
integrated Comptonization parameter 
function YF ({\it left panel}) and maximum Comptonization parameter function yF 
({\it right panel}) histograms for $z=0$ ({\it top, black}), 
$z=0.5$ ({\it middle, red}), 
and $z=1$ ({\it bottom, blue}) in the flat universe.
The smooth curves are the analytic PS predictions at each redshift 
given by equation~(\ref{eq:nps}).
The residual plots give the difference in the logs between the boosted
and nonboosted YFs and yFs.
Note the significant difference in scales of the residuals between the
YFs and yFs.
\label{fig:bnbyF}}
\end{figure}

\appendix

\section{Fitting Simulation Data} \label{sec:appendix}

We use the same basic forms and procedures to fit the merger boosts
discussed in \citetalias{RSR02}, Appendix B.
For the integrated Comptonization parameter, the boosted part of the
cumulative time distribution histograms is well-fit by hyperbolas similar
in form to equation~(B1) of \citetalias{RSR02} with a slight modification:
  \begin{equation}
\ln \left(\frac{t}{t_{\rm sc}}\right)=\sqrt{\left(\bigg\{\frac{Y}{Y(0)}
    -\left[\frac{Y}{Y(0)}\right]_{\rm peak}-1\bigg\}^2-1\right)
    \left(\epsilon^2-1\right)}- \ln \left(\frac{t}{t_{\rm sc}}\right)_Y
    \, .
  \end{equation}
Three parameters describe the function: the maximum boost
$[Y/Y(0)]_{\rm peak}$, the boost duration $(t/t_{\rm sc})_Y$, and the
eccentricity of the hyperbola $\epsilon$.
The fit values for these parameters between simulation runs could be
reproduced with the same functions of fractional mass increase $f_M$
and normalized impact parameter $b'$ used in \citetalias{RSR02},
provided here for completeness:
  \begin{equation}
    \left[\frac{Y}{Y(0)}\right]_{\rm peak}(f_M, b')=\frac{Af_M^B}{C+b'^2}+1
    \, ,
  \end{equation}
  \begin{equation}
    \epsilon(f_M, b')=\left(\frac{Af_M^B}{C+b'^2}\right)
    \, ,
  \end{equation}
  \begin{equation}
    \ln \left(\frac{t}{t_{\rm sc}}\right)_Y=
    G \, \frac{ \ln (M_<+M_>) - H \, \ln (M_<^{1/3}+M_>^{1/3})}
      {I+b'^2}
    \, .
  \end{equation}
As in the text, the impact parameter is scaled by the core radii of the
two merging clusters, $b'=b/(r_{c <} + r_{c >})$, $M_<$ and $M_>$
are the masses of the less massive and more massive cluster
(in $M_{\odot}$), respectively, and the fractional mass increase
$f_M\equiv M_</(M_<+M_>)$.
Motivations for these forms are given in Appendix B of \citetalias{RSR02}.

The variation of $y_{\rm max}$ with the viewing angle of the merger
causes the histograms of values of time versus $y_{\rm max}$ to
be broader than the histograms for $Y$
(Figure~\ref{fig:histy}).
This difference makes hyperbolae a poor representation of the histogram
shapes.
We find a suitable replacement in another 3 parameter function
  \begin{equation}
    \ln \left(\frac{t}{t_{\rm sc}}\right) = P \, \ln
    \left(1-\frac{y_{\rm max}}{y_{\rm peak}}\right)
    -\frac{1}{2}\frac{y_{\rm max}}{y_{\rm peak}}
    - \ln \left(\frac{t}{t_{\rm sc}}\right)_y
    \, ,
  \end{equation}
with similarly defined parameters for the maximum $y_{\rm max}$ boost
$\frac{y_{\rm peak}}{y_{\rm max}(0)}$, the power law slope $P$, and
the boost duration $\ln \left(\frac{t}{t_{\rm sc}}\right)_y$:
  \begin{equation}
    \frac{y_{\rm peak}}{y_{\rm max}(0)}=\frac{Af_M^B}{C+b'^{1.3}}+1
    \, ,
  \end{equation}
  \begin{equation}
    P=\left(\frac{D}{F+b'^{1.5}}\right)^{-1}
    \, ,
  \end{equation}
  \begin{equation}
    \ln \left(\frac{t}{t_{\rm sc}}\right)_y =
    G \, \frac{\ln (M_<+M_>) - H \, \ln (M_<^{1/3}+M_>^{1/3})}
      {I+b'^2}
    \, .
  \end{equation}
The best-fit values found for $A$--$I$ are given in Table~\ref{tab:appendix}.
Note that $A$--$I$ are found assuming that the value of $y_{\rm max}(0)$
is taken along the merger axis, which is twice
the value of $y_{\rm max}(0)$ used in Figures~\ref{fig:yt} and \ref{fig:histy},
for which the value perpendicular to the merger axis is used.

\begin{deluxetable}{lccccccccc}
\tablecaption{Fitting Parameters for Merger Boost Histograms 
\label{tab:appendix}}
\tablewidth{0pt}
\tablehead{Boost & A & B & C & D & E & F & G & H & I}
\startdata
$Y/Y(0)$ & 95.69 & 0.8793 & 66.72 & 94.83 & 0.3621 & 173.3 & 33.36 & 0.2793 & 473.3 \\
$y_{\rm max}/y_{\rm max}(0)$ & 26.55 & 0.5776 & 4.052 & 6.310 & - & 4.569 & 2.250 & 1.785 & 13.76 \\
\enddata
\end{deluxetable}

\end{document}